\documentclass[acmsmall]{acmart}

\usepackage{mdframed}
\usepackage{graphicx}
\usepackage{listings}
\usepackage{subcaption}
\usepackage{tikz}
\usepackage{enumitem}

\definecolor{webblue}{rgb}{0,0,.7}
\definecolor{webgreen}{rgb}{0,.5,0}
\definecolor{webbrown}{rgb}{.6,0,0}

\usepackage{tikz}
\usepackage{makecell}
\usepackage{multirow}
\usepackage{threeparttable} 
\usepackage{booktabs}
\usepackage{diagbox}
\usepackage{amsfonts}
\usepackage{multicol}
\usetikzlibrary{shadows}
\tikzset{every shadow/.style={opacity=1}}
\newmdenv[shadow=false,shadowcolor=black,shadowsize=0pt,linewidth=1pt,skipabove=0pt]{highlightbox}

\newcommand{\textalign}[1]{
    \textcolor{webbrown}{\texttt{#1}}
}

\newcommand{\alldatabar}{%
    \begin{tikzpicture}[scale=0.25, line width=0.1mm]
        \fill[gray] (0,0) rectangle (5, 1);
        \draw (0,0) rectangle (5,1);
    \end{tikzpicture}%
}

\newcommand{\sampledatabar}{%
    \begin{tikzpicture}[scale=0.25, line width=0.1mm]

    \foreach \i in {1, ..., 29} {
        \fill[gray] (\i/6, 0) rectangle (\i/6 + 0.08, 1); 
    }
        \draw (0,0) rectangle (5,1);
    \end{tikzpicture}%
}

\newcommand{\fewdatabar}{%
    \begin{tikzpicture}[scale=0.25, line width=0.1mm]
        \fill[gray] (0,0) rectangle (0.1, 1);
        \fill[white] (0.2,0) rectangle (5, 1);
        \draw (0,0) rectangle (5,1);
    \end{tikzpicture}
}

\lstdefinestyle{SQLStyle}{
  language=SQL,
  basicstyle={\footnotesize\ttfamily},
  breaklines=true,
  frame=single, 
  frameround=tttt, 
  backgroundcolor=\color{gray!5}, 
  rulesep=0pt,
 float=t,
 floatplacement=tbp,  
  numbers=none,
  keepspaces=true,
  showstringspaces=false,
  captionpos=b,
  aboveskip=1pt,
  belowskip=-20pt,
  numberstyle=\tiny\color{gray},  
  stringstyle=\color{webgreen},
  keywordstyle=\color{webblue},
  commentstyle=\color{gray},
  keywords=[2]{LATERAL, UNNEST, APPLY},
  keywordstyle=[2]\color{webblue},
  keywords=[3]{LLM, ObjColumn, ObjTable, SampleBatch, SampleNum, TableName, LimitedNum, ColumnName, ColumnType, ColumnConstraints, ColumnComment, DBName, ColumnSize},
  keywordstyle=[3]\color{webbrown},
  keywords=[4]{RAND},
  keywordstyle=[4]\color{webgreen},
}

\lstdefinestyle{prompt}{
    backgroundcolor=\color{gray!5},
    basicstyle={\footnotesize\ttfamily},
    keepspaces=true,
    showstringspaces=false,
    numbers=none,
    frame=single,
    xleftmargin=3pt,
    xrightmargin=3pt,
    aboveskip=5pt,
    belowskip=-10pt,
    captionpos=b,
    breaklines=true,             
    breakindent=-1pt,
    breakatwhitespace=true,
    tabsize=2,
    keywords=[3]{ColumnName, ColumnType, ColumnConstraints, ColumnComment, ColumnSize, },
  keywordstyle=[3]\color{webbrown},
  keywords=[4]{Samples, TableInfo},
  keywordstyle=[4]\color{webgreen},
  morecomment = [s][\color{webgreen}\bfseries]{Data samples of the current }{column}, 
}

\lstdefinestyle{prompt2}{
    backgroundcolor=\color{gray!5},
    basicstyle={\footnotesize\ttfamily},
    keepspaces=true,
    showstringspaces=false,
    numbers=none,
    frame=single,
    xleftmargin=3pt,
    xrightmargin=3pt,
    aboveskip=1pt,
    belowskip=1pt,
    captionpos=b,
    breaklines=true,            
    breakindent=-1pt,
    breakatwhitespace=true,
    tabsize=2,
    keywords=[3]{ColumnName, ColumnType, ColumnConstraints, ColumnComment, ColumnSize, k,ndv_first_k_values,min_first_k_values,max_first_k_values, most_frequent_first_k_values,first_k_values },
  keywordstyle=[3]\color{webbrown},
  keywords=[4]{Samples, TableInfo},
  keywordstyle=[4]\color{webgreen},
  morecomment = [s][\color{webgreen}\bfseries]{Data samples of the current }{column}, 
}

\lstdefinelanguage{Prompt}{
	backgroundcolor=\color{backcolour},   
	keywordstyle=\color{magenta},
	numberstyle=\tiny\color{codegray},
	basicstyle=\ttfamily\footnotesize,
	breakatwhitespace=false,         
	breaklines=true,   
    breakindent=-3.5pt,
	captionpos=b,                    
	keepspaces=true,                 
	numbers=left,                    
	numbersep=5pt,                  
	showspaces=false,                
	showstringspaces=false,
	showtabs=false,                  
	tabsize=4,
	escapeinside={`}{`},
	morecomment = [s][\color{eclipseGreen}\bfseries]{How}{?},
        morecomment = [l][\color{eclipseBlue}\bfseries]{SELECT},
        morecomment = [l][\color{eclipsePurple}\bfseries]{\$\{DATABASE_SCHEMA\}},
        morecomment = [s][\color{eclipsePurple}\bfseries]{CREATE}{;},
        morecomment = [l][\color{eclipsePurple}\bfseries]{Table},
        morecomment = [l][\color{eclipsePurple}\bfseries]{continents},
        morecomment = [l][\color{eclipsePurple}\bfseries]{countries},
        morecomment = [l][\color{codewhite}\bfseries]{\$\{TARGET_QUESTION\}},
    postbreak={
       \mbox{
           \lst@linebreakbgrd
           \rotatebox[y=0.7ex]{180}{\color{black}$\Lsh\,$}
       }
    },
}
\definecolor{pythonblue}{RGB}{0, 0, 255}
\definecolor{pythongreen}{RGB}{0, 128, 0}
\definecolor{pythonred}{RGB}{255, 0, 0}
\definecolor{pythonorange}{RGB}{255, 165, 0}
\definecolor{pythonpurple}{RGB}{128, 0, 128}

\lstdefinestyle{python}{
    language=Python,
    basicstyle={\footnotesize\ttfamily},
    backgroundcolor=\color{gray!5},  
    keywordstyle=\color{blue},      
    commentstyle=\color{pythongreen},       
    stringstyle=\color{red},         
    numbers=left,                    
    numberstyle=\tiny\color{gray},    
    stepnumber=1,                   
    numbersep=5pt,                   
    showspaces=false,                 
    showstringspaces=false,         
    showtabs=false,                   
    tabsize=4,                       
    frame=single,                      
    breaklines=true,  
    captionpos=b,   
    xleftmargin=3pt,
    xrightmargin=3pt,
}

\setcopyright{acmlicensed} 
\acmJournal{PACMMOD}
\acmYear{2025} \acmVolume{3} \acmNumber{3 (SIGMOD)}
\acmArticle{199} \acmMonth{6} 
\acmDOI{10.1145/3725336}
\begin{document}

\title{PLM4NDV: Minimizing Data Access for Number of Distinct Values Estimation with Pre-trained Language Models}


\author{Xianghong Xu}\orcid{0000-0003-2447-4107}
\affiliation{
  \institution{ByteDance}
  \city{Beijing}
  \country{China}
}
\email{xuxianghong@bytedance.com}

\author{Xiao He}\orcid{0000-0001-7959-2157}
\affiliation{
  \institution{ByteDance}
  \city{Hangzhou}
  \country{China}
}
\email{xiao.hx@bytedance.com}

\author{Tieying Zhang}\orcid{0009-0003-2250-5528}
\affiliation{
  \institution{ByteDance}
  \city{San Jose}
  \country{USA}
}
\email{tieying.zhang@bytedance.com}\authornote{Tieying Zhang corresponds to this work.}

\author{Lei Zhang}\orcid{0009-0004-1681-1956}
\affiliation{
  \institution{ByteDance}
  \city{San Jose}
  \country{USA}
}
\email{zhanglei.michael@bytedance.com}

\author{Rui Shi}\orcid{0009-0003-9122-4703}
\affiliation{
  \institution{ByteDance}
  \city{Beijing}
  \country{China}
}
\email{shirui@bytedance.com}

\author{Jianjun Chen}\orcid{0000-0002-3734-892X}
\affiliation{
  \institution{ByteDance}
  \city{San Jose}
  \country{USA}
}
\email{jianjun.chen@bytedance.com}


\begin{abstract}
Number of Distinct Values (NDV) estimation of a multiset/column is a basis for many data management tasks, especially within databases. Despite decades of research, most existing methods require either a significant amount of samples through uniform random sampling or access to the entire column to produce estimates, leading to substantial data access costs and potentially ineffective estimations in scenarios with limited data access. In this paper, we propose leveraging semantic information, i.e., schema, to address these challenges. The schema contains rich semantic information that can benefit the NDV estimation. To this end, we propose \texttt{PLM4NDV}, a learned method incorporating Pre-trained Language Models (PLMs) to extract semantic schema information for NDV estimation. Specifically, \texttt{PLM4NDV} leverages the semantics of the target column and the corresponding table to gain a comprehensive understanding of the column's meaning. By using the semantics, \texttt{PLM4NDV} reduces data access costs, provides accurate NDV estimation, and can even operate effectively without any data access. Extensive experiments on a large-scale real-world dataset demonstrate the superiority of \texttt{PLM4NDV} over baseline methods. Our code is available at \url{https://github.com/bytedance/plm4ndv}.
\end{abstract}

\ccsdesc[500]{Information systems~Query optimization}

\keywords{Number of Distinct Values; Minimize Data Access; Pre-trained Language Model}

\received{October 2024}
\received[revised]{January 2025}
\received[accepted]{February 2025}

\maketitle

\section{Introduction}\label{sec:intro}
Estimating the number of distinct values (NDV) of a column is a fundamental problem in various domains, including databases~\cite{hybskew_haas1995sampling,brutlag2002block}, networks~\cite{network_cohen2019cardinality,network_nath2008synopsis}, biology~\cite{valiant2013estimating,valiant2017estimating,mmo_bunge1993estimating} and statistics~\cite{goodman1949estimation,chao1984nonparametric}. 
In the field of databases, NDV estimation is the foundation of query optimization. 
For instance, the relative error in the join-selectivity formulas used in MySQL~\cite{mysql_join} is directly related to the relative error in the constituent NDV estimations~\cite{hybskew_haas1995sampling}. 
Moreover, systems like Spark~\cite{spark_plan_code} and PostgreSQL~\cite{pg_plan_code} rely on NDV to compute cardinality, a crucial metric in query optimization. 
Additionally, recent studies also demonstrate that precise NDV estimation can lead to more efficient query plans that significantly reduce the execution latency for different database management systems (DBMSs)~\cite{li2023alece,han2024bytecard}.
Furthermore, the effectiveness of some external index advisors is strongly influenced by the accuracy of NDV estimations~\cite{lib_shi2022learned,idxl_peng2023data,idxl_peng2024online,yu2024refactoring}.

Approaches for estimating NDV without an index can be broadly categorized into two classes: \textit{sketch-based} and \textit{sampling-based}. 
The key difference between the two categories lies in whether the full data is accessed. 
The sketch-based methods require scanning the entire table once and creating memory-efficient sketches~\cite{histogram_selinger1979access,flajolet2007hyperloglog}, which are then used to estimate NDV. 
Sampling-based methods, on the other hand, access a small randomly selected subset of the data and use the statistical features derived from these samples to estimate NDV~\cite{hybskew_haas1995sampling,li2024learning,ls_wu2022learning}.
Nevertheless, despite the significant progress achieved, existing methods still incur notable costs of data access when it comes to NDV estimation for in-production DBMSs.

Sketch-based methods are the most accurate NDV estimators; however, the full data access requirement leads to high I/O costs. 
For a production database with an OLTP workload running, the cost of accessing the full data is prohibitive. 
In such situations, sampling-based methods are the only alternatives, which seem to fit the situation perfectly at first glance.
Nevertheless, they still incur significant data access costs when applied to production DBMSs, even though they only need a small subset of samples.
The primary reason is that almost all sampling-based methods~\cite{goodman1949estimation,chao1984nonparametric,chao_in_db_ozsoyoglu1991estimating,chaolee,shlosser1981estimation,horvitz_sarndal1992model} require the samples to be independent and identically distributed (IID), leading to sampling uniformly at random~\cite{olken1993random}. 
It is difficult to implement efficiently in DBMSs due to high random I/O costs on disk. Therefore, many applications rely on running SQL statements to obtain samples from the DBMS~\cite{DBLP:conf/icde/HeT00ZLX23,han2024bytecard,idxl_peng2024online,idxl_peng2023data} with a strict sampling cost budget. 
Specifically, sampling a small portion of data uniformly at random using standard SQL from a table with millions of rows can take more than ten seconds in practice. Besides, random sampling may lead to database failures when the database instance has a heavy workload. 
Though block-level sampling methods~\cite{brutlag2002block,chaudhuri2004effective} have been developed to obtain samples at the block (or page) level in some commercial DBMSs to improve efficiency, a considerable number of blocks or pages still need to be accessed.

\noindent\textbf{Motivation.} 
Motivated by the limitations of existing studies and the escalating practical requirements of data access cost constraints for NDV estimation in real-world applications, this paper endeavors to explore the feasibility of achieving precise NDV estimation with minimal sequential data access or even without any data access. 
Sequentially accessing the top hundred records violates the IID sample requirement of sampling-based methods but offers greater efficiency by reducing I/O costs. Moreover, operating without any data access eliminates data retrieval costs entirely, yet this approach has not been thoroughly investigated.
To address these challenges, we propose leveraging database schemas to achieve precise NDV estimation with highly restricted data access. The semantics in schemas are closely related to the NDV of specific columns, making them a valuable resource for this task.

\begin{lstlisting}[style=SQLStyle, label=lst:intro:schema, caption={An illustration of a schema in MySQL.}]
CREATE TABLE TableName(
ColumnName ColumnType ColumnConstraints, ColumnComment
EmployeeID int NOT NULL, COMMENT 'Identifier for each employee, unique in this table',
EmployeeNation VARCHAR(30) NOT NULL, COMMENT 'Nationality for each employee',
-- Definitions of indexes below are omitted.
);
\end{lstlisting}

In the following, an example schema with a MySQL database is presented in Listing~\ref{lst:intro:schema} to illustrate our proposed concept. Herein, \lstinline[style=SQLStyle]{TableName} represents the name of the table. Meanwhile, the \lstinline[style=SQLStyle]{ColumnName}, \lstinline[style=SQLStyle]{ColumnType}, and \lstinline[style=SQLStyle]{ColumnConstraints} define the name, data type, and data constraints of a column, respectively. Additionally, the \lstinline[style=SQLStyle]{ColumnComment} is the comments usually provided by database administrators (DBA) to describe the structures, meanings, and usages of each column. To exemplify this, consider the columns \lstinline[style=SQLStyle]{EmployeeID} and \lstinline[style=SQLStyle]{EmployeeNation}, where the semantic information provides useful insights corresponding to their NDVs. 
From the column name and comments, it is evident that \lstinline[style=SQLStyle]{EmployeeID} serves as a unique identifier for employees, implying that its NDV is likely to be comparable to the number of records within the table.
Conversely, \lstinline[style=SQLStyle]{EmployeeNation} would have an NDV that does not exceed the total number of recognized countries worldwide. Consequently, leveraging such semantic information can significantly enhance the accuracy of NDV estimations, thereby substantially reducing data access.

\noindent\textbf{Our approach.} 
Inspired by the aforementioned observations, this paper introduces \texttt{PLM4NDV}, a novel approach that integrates semantic information utilizing pre-trained language models (PLMs) for NDV estimation. PLMs such as BERT~\cite{kenton2019bert}, GPT~\cite{gpt_radford2018improving}, and GPT2~\cite{gpt2_radford2019language} have demonstrated state-of-the-art (SOTA) performance across a wide range of natural language processing (NLP) tasks. Consequently, we propose leveraging a PLM to extract semantic features from the target column for NDV estimation.
The design of \texttt{PLM4NDV} adheres to the following principles:
(1) \textbf{Minimize data access.} \texttt{PLM4NDV} targets the practical NDV estimation problem with only limited sequential data access, e.g., accessing only the first hundred records.
(2) \textbf{Semantics are critical.} The schema is indispensable for the storage of tabular data within a DBMS. Using semantics in the schema may be beneficial for NDV estimation, therefore, incorporating the output of PLM with a learned model can enhance NDV estimation.
(3) \textbf{Auxiliary of table.} Given that \texttt{PLM4NDV} relies on semantics, we consider the auxiliary semantics of columns within the same table to estimate the NDV of the target column. We believe that the information in the table can improve the estimation accuracy of individual columns.

\begin{table}
  \caption{Comparison of NDV estimation approaches, where the gray shadows indicate the accessed data and the shaded area roughly represents the relative size.}
  \label{tab:methodcategories}
  \begin{tabular}{cccc}
    \toprule
    Categories & Data Access  & Method Property\\
\midrule
    sketch & \alldatabar  &  Memory-efficient sketches \\
\midrule
    \makecell{sampling (learning)} & \sampledatabar & \makecell{Statistics of IID samples} \\
\midrule
    semantic & \fewdatabar  & \makecell{Semantics with optional statistics} \\
  \bottomrule
\end{tabular}
\end{table}

Existing methods are inapplicable without data access, whereas \texttt{PLM4NDV} can estimate NDV even in the absence of data, effectively addressing this gap. Scenarios where data access is unavailable do occur in practice, such as when users set access permissions due to privacy concerns or when the sampling cost budget is so limited that not even a single sample can be obtained. 
Although \texttt{PLM4NDV} may not always achieve very high accuracy using solely semantic information, it provides a unique option to address the issue.
Therefore, \texttt{PLM4NDV} is distinctly different from existing methods and does not belong to sketch-based or sampling-based categories.
Recent learning-based methods~\cite{ls_wu2022learning, li2024learning} are categorized as sampling-based because they rely on IID samples as input.
In contrast, our method represents a new category of NDV estimation: semantic-based. This distinction is illustrated in Table~\ref{tab:methodcategories}.
We hope that our approach will inspire the community to explore and develop more semantic-based NDV estimation methods and other statistical estimation techniques that effectively leverage semantic information.

\noindent\textbf{Our contributions.} This paper makes the following contributions.

\begin{itemize}
    \item We propose \texttt{PLM4NDV}, a practical NDV estimation method that minimizes the data access by incorporating PLM to extract the semantic features of columns. To the best of our knowledge, this is the first approach to leverage semantic information for NDV estimation.
    \item \texttt{PLM4NDV} provides the only alternative approach when the data access cost budget restricts sampling any data. We introduce a new paradigm in NDV estimation with minimal data access costs: a semantic-based approach that is distinct from both sketch-based and sampling-based methods.    
    \item Our extensive experiments on a large-scale real-world dataset demonstrate the promising performance of \texttt{PLM4NDV}, particularly in scenarios where data access is limited. The experiments also reveal several insightful findings that inform future research directions.
\end{itemize}


\section{Preliminaries}\label{sec:preliminary}
\noindent\textbf{Problem statement.} Consider a column $C=(c_1,c_2,\ldots,c_N)$ consisting of $N$ items where each item $c_i(1\leq i\leq N,c_i\in\Omega)$ is a member of a universe of $D(D\leq N)$ possible values. Different items within the column may share the same value. Let $N$ denote the size of the column $C$ and let $D$ represent the NDV within $C$. A subset $S$ of $C$ is sampled either randomly or sequentially, with $n$ denoting the size of $S$. The NDV of $S$ is represented by $d$ and the sampling rate is given by $r=n/N$. 
The column $C$ has a name and data type defined within the database schema. Additionally, data constraints and comment descriptions of the column may be present in the schema.
The objective is to estimate $D$ utilizing the schema information and the sample $S$.

\noindent\textbf{Semantics.} Semantics refer to the meaning or interpretation of the data within the column, expressed in natural language. We assume that the semantics of the stored data align with the column name defined in the schema and a column with a meaningful column name tends to have certain NDV distributions, suggesting that leveraging semantics may be beneficial for NDV estimation. Misalignment can occur in real-world scenarios, for instance, a column storing dates may be arbitrarily named \lstinline[style=SQLStyle]{ColumnA} by users. However, advanced techniques in the column annotation task~\cite{suhara2022annotating} can effectively address these issues. Consequently, our assumptions are reasonable in practice.

\noindent\textbf{Sample Statistics.} Two key sample statistics are widely used in NDV estimation, and they are defined as follows: 

\textit{Frequency.} The frequency of a value $c_i$ in a column is the number of times it appears. Let $N_{c_i}$ denote the frequency of $c_i$ in $C$ and $n_{c_i}$ denotes the frequency of $c_i$ in the sample $S$. By definition, $n=\sum_{c_i}n_{c_i}$ and $d=|\{c_i\in\Omega|n_{c_i}>0\}|$. 

\textit{Frequency profile.} The frequency profile represents the frequency of frequencies. Let $F_j=|\{i\in\Omega|N_{c_i}=j\}|$ denote the NDV with frequency $j$ in $C$, and $f_j=|\{i\in\Omega|n_{c_i}=j\}|$ denote the NDV with frequency $j$ in $S$. By this definition, $n=\sum_j j\cdot f_j$ and $d=\sum_{j} f_j$.

\noindent\textbf{Illustration of NDV estimation methods.} We demonstrate how various methods establish NDV estimation, excluding sketch-based methods, which are not applicable under limited data access. 

\textit{Sampling-based methods.} Sampling-based methods primarily leverage the statistics of the IID samples to derive the estimation. In general, a formal description is given by $\hat{D}=\mathcal{M}(N,f)$, where $\hat{D}$ denotes the estimated NDV and $\mathcal{M}$ represents the general estimation function. For instance, a learned NDV estimation method~\cite{ls_wu2022learning} takes the cut-off frequency profiles and $N$ as input and estimates NDV by a learnable multi-layer perception (MLP). The \texttt{Sichel}~\cite{sichel1986parameter,sichel1986word,sichel1992anatomy} method requires solving non-linear equations:
\begin{align}
    \begin{aligned}
        (1+g)\ln g-Ag+B=0,\frac{f_1}{n}<g<1, A=\frac{2n}{d}-\ln\frac{n}{f_1},\\
        B=\frac{2f_1}{d}+\ln \frac{n}{f_1},\hat{b}=\frac{g\ln \frac{ng}{f_1}}{1-g},\hat{c}=\frac{1-g^2}{ng^2},
        \hat{D}_{\mathrm{\texttt{Sichel}}}=\frac{2}{\hat{b}\hat{c}}.\\
    \end{aligned}
    \label{eq:sichel}
\end{align}

Another representative method, \texttt{Goodman}~\cite{goodman1949estimation}, constructs a linear polynomial using $N$ and $f$ to establish the estimation $\hat{D}_{\mathrm{\texttt{Goodman}}}$:
\begin{equation}
\hat{D}_{\mathrm{\texttt{Goodman}}} =d+\sum_{i=1}^n(-1)^{i+1} \frac{(N-n+i-1) !(n-i) !}{(N-n-1) ! n !} f_i.
    \label{eq:goodman}
\end{equation}

\textit{Semantic-based methods (ours).} Semantic-based methods utilize the semantics in the schema along with optional sample statistics. The formal description is $\hat{D}=\mathcal{M}(semantics,N,\underline{f})$, where $\mathcal{M}$ is the general function and $\underline{f}$ represents $f$ is optional in the formula.


\section{Methodology of \texttt{PLM4NDV}}

\begin{figure*}
    \centering
    \includegraphics[width=\linewidth]{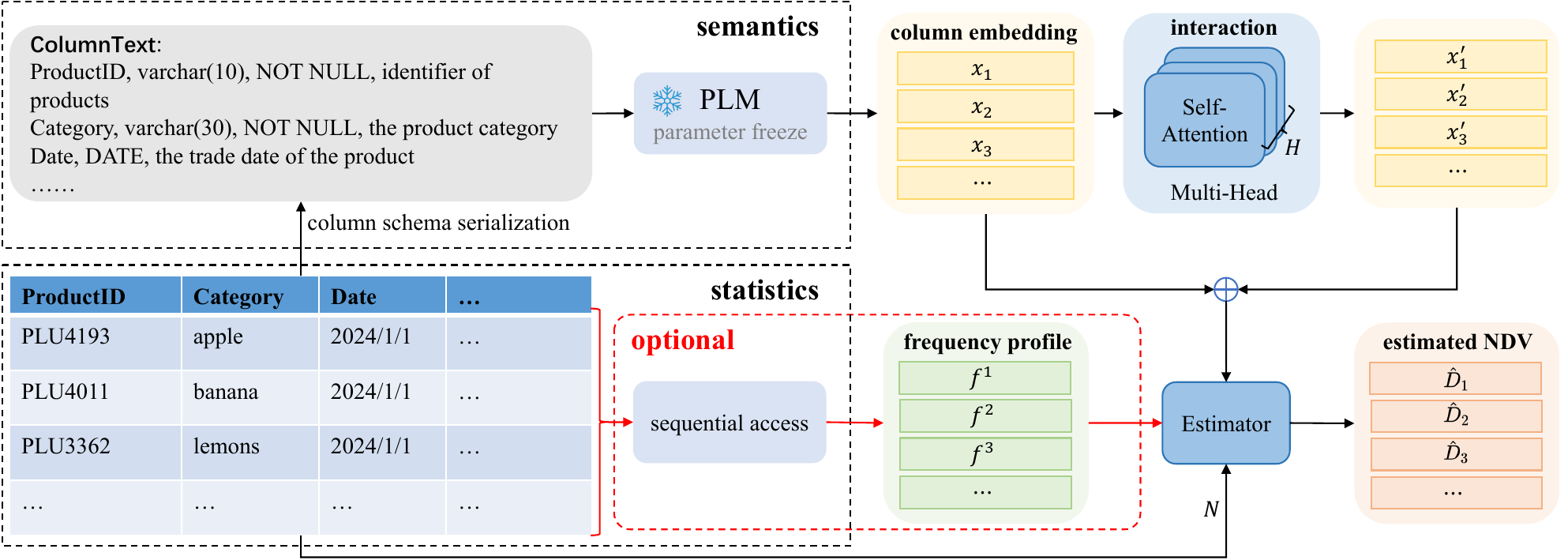}
    \caption{The model architecture of \texttt{PLM4NDV} is illustrated using a hypothetical table. The statistic component is akin to existing sampling-based methods but operates more efficiently and with a reduced volume of data. In addition, accessing data in the table is optional for \texttt{PLM4NDV}.}
    \label{fig:plm4ndv}
\end{figure*}

\subsection{Model Architecture}
The architecture of \texttt{PLM4NDV} is depicted in Figure~\ref{fig:plm4ndv}, utilizing a hypothetical table that stores data related to various fruits to illustrate the model's estimation process. Our model relies on two main categories of features: semantic information and sample statistics. We consider accessing the first page of data as the limited data scenario because it incurs minimal costs when accessing data. In many database systems, a single page usually contains dozens to hundreds of rows, depending on the specific database and the width of the table. In this paper, we use the first 100 rows as an example to illustrate the limited data scenario.

\noindent\textbf{Data pipelines.} The semantic features encompass characteristics related to both individual columns and the corresponding table. Initially, the schema is serialized into natural language texts for each column. These textual representations are then processed through a PLM to derive the embeddings for each column. Subsequently, the column embeddings within the same table are fed into the column interaction component, thereby generating table-aware representations. Statistical features consist of the frequency profiles of each column derived from sequentially accessed samples, which are optional inputs for the \texttt{PLM4NDV} estimation process. The learned estimator utilizes both semantic and statistical inputs to predict the NDVs for each column.

\noindent\textbf{Training.} During the training phase, we gather a large set of tables from the real world and record the following information for each column: the schema, the statistics of the top 100 rows, and the ground-truth NDV. By this means, our method can learn the relationship between the schema semantics and the corresponding NDV in real-world applications. The parameters of the PLM are kept frozen, therefore, the training of \texttt{PLM4NDV} focuses solely on the column interaction and NDV estimator modules.

\noindent\textbf{Inference.} Following training on existing datasets, \texttt{PLM4NDV} can perform estimations on unseen tables. First, the same PLM is used to extract semantic representations from the schema of the test table. Next, sample statistics are gathered from the first 100 rows of each column, which are sequentially accessed. Finally, the trained model is employed to estimate the NDV.

\subsection{Column Embedding}
\label{ref:columnembedding}
As shown in Figure~\ref{fig:plm4ndv}, the first step of \texttt{PLM4NDV} involves leveraging the structured schema by serializing it into natural language. Then, we extract semantic features from the serialized natural language text.

\noindent\textbf{Schema serialization.} Since PLMs require token sequences (i.e., natural language text) as input, the initial step involves converting the database schema into text sequences for each column. This transformation enables the structured schema to be processed by PLMs. Specifically, the serialized text sequence of a column is defined as follows:
\begin{center}
\texttt{ColumnText}:=\textalign{ColumnName},\textalign{ColumnType},\textalign{ColumnConstraints},\textalign{ColumnComment}
\end{center}
where the column definitions and descriptions are concatenated using commas, employing a straightforward way to transform the schema of a column into a sequence of tokens.

In addition, a substantial number of column schemas may lack the components \lstinline[style=SQLStyle]{ColumnConstraints} and \lstinline[style=SQLStyle]{ColumnComment} in practical scenarios. We assume \lstinline[style=SQLStyle]{ColumnName} and \lstinline[style=SQLStyle]{ColumnType} are available for the model input, while \lstinline[style=SQLStyle]{ColumnConstraints} and \lstinline[style=SQLStyle]{ColumnComment} are optional. The example of the \lstinline[style=SQLStyle]{Date} column in Figure~\ref{fig:plm4ndv} illustrates this situation. For columns that are missing these optional components in the schema, we exclude these elements and concatenate the serialized available components.

\noindent\textbf{Semantic embedding.} We leverage sentence transformers~\cite{reimers2019sentence,ni2021sentencet5} as our foundational models due to their proficiency in producing similar vector-space representations for semantically analogous sentences. This capability is particularly suitable for extracting semantic information for columns in databases. By leveraging these models, we can effectively capture the semantic relationships between different columns. For instance, the \lstinline[style=SQLStyle]{ProductID}, \lstinline[style=SQLStyle]{ProductCode}, and \lstinline[style=SQLStyle]{ProductNumber} columns may possess different names (and thus different serialized text sequences) but convey similar semantic meanings. The PLMs enable us to process these variable-length text sequences as input and generate fixed-size vectors, thereby facilitating the utilization of schema information.

Since we require only the semantic embeddings of the serialized column text from the PLM, we can utilize the frozen parameters of the PLM to obtain the column embeddings without fine-tuning. Specifically, the column embedding can be formulated as: $x=\texttt{PLM}(\texttt{ColumnText}),x\in\mathbb{R}^l$, where $l$ is the embedding size of the PLMs.

\subsection{Column Interaction}\label{sec:column-interaction}
Columns with the same semantics may exhibit different selectivities in varying table contexts. Therefore, the semantics from other columns within the same table can provide valuable auxiliaries in NDV estimation for individual columns. For instance, consider a table \lstinline[style=SQLStyle]{Product} containing two columns named \lstinline[style=SQLStyle]{State} and \lstinline[style=SQLStyle]{FIPSCode} (Federal Information Processing Standards Code used in the USA), which store relevant information of products, respectively.
Estimating the NDV of \lstinline[style=SQLStyle]{State} alone using only semantic information may be challenging, as it is unclear whether the states refer to those in the USA or other countries. The \lstinline[style=SQLStyle]{FIPSCode} column provides valuable auxiliaries, indicating that the \lstinline[style=SQLStyle]{State} likely refers specifically to states within the USA.

Given this insight, we believe that the auxiliary semantic information from other columns within the same table can be beneficial. Therefore, we take the embeddings of all columns in a table as input to establish the interactions between them. Specifically, let the column embeddings in a table be denoted as:
\begin{align}
X=[x_1,x_2,\ldots,x_t]^\top ,x_i\in\mathbb{R}^{l},
\end{align}
where there are $t$ columns in the table and $x_i$ is the embedding of the $i$-th column from a frozen PLM. For each column embedding $x_i$, we consider the information from other columns, thereby enhancing the overall understanding of the semantics of individual columns within the table. To this end, we leverage Multi-Head Self-Attention (MHSA)~\cite{vaswani2017attention} to achieve our goals as illustrated in Figure~\ref{fig:plm4ndv}. The self-attention mechanism enables the model to focus on relevant aspects of the input data while disregarding irrelevant information, thereby facilitating the learning of relationships between the columns in the table.

Particularly, the attention mechanism involves three components: Query ($Q$), Key ($K$), and Value ($V$), which are derived through three linear transformations:
\begin{align}
    Q=XW^Q+b^Q,K=XW^K+b^K,V=XW^V+b^V,
    \label{eq:qkv}
\end{align}
where $W^Q,W^K,W^V\in\mathbb{R}^{l\times l}$ are learnable weight matrices and $b^Q,b^K,b^V\in\mathbb{R}^l$ are learnable bias vectors. The Self-Attention (SA) mechanism is formulated as:
\begin{align}
    SA(Q,K,V)=\text{softmax}\left(\frac{QK^\top}{\sqrt{l}}\right)V,
\end{align}
where softmax~\cite{bridle1990probabilistic} converts the dot-product correlations between the elements in $Q$ and $K$ into normalized attention coefficients (probability distribution). This allows for a weighted sum to be applied to $V$, thereby highlighting the relevant information from the table. The MHSA can be obtained by:
\begin{align}
    MHSA(Q,K,V)=[SA_1,SA_2,\ldots,SA_H]W + b,
\end{align}
where $H$ is the number of heads, and each SA component undergoes individual transformations as defined in Eqition~(\ref{eq:qkv}). The feature dimension for each SA head is $l^\prime=H\cdot l$. The weight matrix $W\in\mathbb{R}^{l^\prime\times l^\prime}$ and the bias vector $b\in\mathbb{R}^{l^\prime}$ are applied after concatenating the outputs from the multiple heads. Consequently, the output dimension of MHSA is denoted as $X^\prime\in\mathbb{R}^{t\times l^\prime}$, where $x^\prime_i\in\mathbb{R}^{l^\prime}$ represents the column-interacted table semantics representation of the $i$-th column. For more details about MHSA not covered in this paper, please refer to~\cite{vaswani2017attention}.

\subsection{Statistics Collection}\label{sec:seqdata}
Although our method can estimate NDV without accessing data, relying solely on semantic information does not yield high estimation accuracy. Due to the diversity in data distributions in real-world applications or the effects of database sharding, two tables with similar or identical definitions may exhibit substantial differences in NDV. Continuing from the previous example of table \lstinline[style=SQLStyle]{Product}, another database may contain the same table but store product content from different regions, or multiple sharded databases may be organized according to specific categories by DBAs. Consequently, it is essential to access a portion of the data for accurate estimations.

\noindent\textbf{Data access.} Random sampling is usually implemented by running SQL queries, which can hardly satisfy a limited sampling cost budget. Therefore, we utilize sequential access to get data samples. We directly access the top $n$ rows (e.g., $n$=100) for each column, which allows us to complete data access with minimal (i.e., once) I/O operation, significantly reducing the access costs.

\noindent\textbf{Frequency profile construction.} 
The accessed data sample is denoted as $S$ and $|S|=n$. Each item $s_i$ in $S$ is a tuple containing $t$ elements, where $t$ is the number of columns. We construct the frequency profiles for each column (illustrated in Section~\ref{sec:preliminary}) in $S$, and $f^i\in\mathbb{R}^n$ represents the frequency profile of the $i$-th column.

\noindent\textbf{Number of rows.} Essentially, the total number of rows $N$ is required for the estimation, and our method also leverages this information. The cost of accessing $N$ is minimal in practice because the DBMS typically maintains these basic statistics. For instance, in MySQL, the number of rows for the tables can be assessed using the SQL query: \lstinline[style=SQLStyle]{SHOW TABLE STATUS FROM DBName;}, where \lstinline[style=SQLStyle]{DBName} is the database name containing the object tables.

\subsection{Learned NDV Estimation}

\noindent\textbf{Learned estimator.} The learned estimator takes both the semantic and statistical features as input to estimate NDV. We use an MLP to map the input features into the estimation:
\begin{align}
    \log\hat{D_i}=MLP([x_i+x^\prime_i,\log N_i,\underline{f^i}]),
\end{align}
where $\hat{D_i}$ is the estimated NDV of the $i$-th column in the dataset, $x_i$ is the semantic feature vector of the column, $x^\prime_i$ is the feature vector of the column that interacted with other columns within the same table, $f^i$ is the frequency profile vector of the data samples, $\underline{f^i}$ represents $f^i$ is optional in the formula, $N_i$ is the size of the column, and $[\cdot]$ indicates feature concatenation. We employ the logarithm operation on $N$ and the estimation $\hat{D}$ to reduce their magnitude and mitigate the impact of large values.

\noindent\textbf{Model learning.} Our objective is to minimize the difference between the estimation and the ground truth, the loss function is formulated as:
\begin{align}
    \mathcal{L}=\frac{1}{\mathcal{N}}\sum_{i=1}^\mathcal{N}(\log \hat{D}_i-\log D_i)^2,
    \label{eq:loss}
\end{align}
where $\mathcal{N}$ is the number of training data points, $\hat{D}_i$ and $D_i$ are the estimation result and the ground truth of the $i$-th sample in the training set.

\section{Experiments}

\subsection{Experimental Setup}\label{sec:setting}

\noindent\textbf{Dataset selection.} Due to the diversity of NDV estimation methods, different approaches focus on various aspects beyond estimation accuracy. Sketch-based methods emphasize memory usage, whereas sampling-based methods consider the sampling size. Therefore, most existing works are evaluated under various specific settings~\cite{review_harmouch2017cardinality} and there is no publicly available, widely used evaluation dataset for NDV estimation. Furthermore, most sampling-based methods are evaluated on columns that follow standard distributions or on several manually crafted datasets that satisfy their assumptions~\cite{goodman1949estimation,motwani2006distinct,mmo_bunge1993estimating,gee_charikar2000towards,sichel1986parameter,sichel1986word,sichel1992anatomy,bootstrap_smith1984nonparametric,hybskew_haas1995sampling}. Recent sampling-based works expand the evaluation to some open-source datasets~\cite{ls_wu2022learning,li2024learning}.

This situation may lead to two primary problems. On the one hand, a limited number of evaluation columns leads to incomplete evaluations. On the other hand, manually crafted columns differ from practical scenarios due to their lack of meaningful schema and monotonous data distribution. Therefore, in this paper, we evaluate our method on a large-scale dataset derived from real-world scenarios. In recent years, the database community has collected extensive relational tabular datasets~\cite{hulsebos2023gittables,eggert2023tablib}. TabLib~\cite{eggert2023tablib} is the largest dataset, containing 627M individual tables totaling 69 TiB. We select the TabLib sample version dataset~\cite{tablib-v1-sample}, which comprises 0.1\% of the full version (69 GB), to evaluate our method. 

\begin{table}[t]
  \caption{Statistics of preprocessed datasets.}
  \label{tab:dataset}
  \begin{tabular}{ccccc}
    \toprule
    & train & test & validation \\
\midrule
    \# Table & 4003 & 1128 & 1267   \\
    \# Column & 34761 & 8725 & 10981    \\
  \bottomrule
\end{tabular}
\end{table}

\noindent\textbf{Data preprocess.} The selected dataset contains 77 parquet files; however, we remove three of them (2d7d54b8, 8e1450ee, and dc0e820c) due to the storage and memory constraints. Each parquet file includes thousands of tables, and we split the remaining parquet files into training, testing, and validation sets. Previous works~\cite{ls_wu2022learning,li2024learning} evaluate NDV estimation methods on large tables (exceeding one million rows) encompassing a total of 218 individual columns. In contrast, we broadened the range of table sizes from tens of thousands to several million rows to provide a more comprehensive evaluation. We filter out columns lacking useful semantics in their names due to misalignment issues, as column annotation~\cite{suhara2022annotating} is beyond the scope of this paper. This includes columns with names consisting of one character or fewer, as well as those composed entirely of numbers, scientific notation, or timestamps.
The statistics of the preprocessed datasets are shown in Table~\ref{tab:dataset}, with the largest table containing 696 columns. As shown in Table~\ref{tab:dataset}, the dataset used in this paper is substantial for evaluation purposes.

We conducted a statistical analysis of the data type of each column in the dataset. We observe that the majority of the data types are primitive, with very few composite types: 34.7\% are \lstinline[style=SQLStyle]{big int}, 33.4\% are \lstinline[style=SQLStyle]{string}, and 30.1\% are \lstinline[style=SQLStyle]{double}. The other data types present in very small quantities include \lstinline[style=SQLStyle]{bool}, \lstinline[style=SQLStyle]{timestamp}, \lstinline[style=SQLStyle]{unsigned int}, \lstinline[style=SQLStyle]{time}, \lstinline[style=SQLStyle]{list}, and \lstinline[style=SQLStyle]{dict}.
The first three common data types account for over 98\% of the dataset, so the experiment analysis in this paper focuses primarily on them.

\noindent\textbf{Evaluation metrics.} Ratio error, known as q-error~\cite{q_error_moerkotte2009preventing}, is widely used to evaluate the performance of NDV estimation:

\begin{equation}
    \mathrm{q\text{-}error}=\max\left(\frac{\hat{D}}{D},\frac{D}{\hat{D}}\right),
    \label{eq:q-error}
\end{equation}
where $\hat{D}$ is the estimated NDV and $D$ is the ground truth NDV. The error is always greater than or equal to 1, and a lower value indicates better performance.

\noindent\textbf{Implementation Details.} The MLP used in \texttt{PLM4NDV} estimator consists of three hidden layers with sizes 384, 128, and 64, respectively. The activation function employed is ReLU. The semantic embedding model is Sentence T5~\cite{ni2021sentencet5}. We train the model by the Adam~\cite{kingma2014adam} optimizer with an initial learning rate of 0.001. The training batch size is 256. Model checkpoints are saved according to the 90\% percentile of q-error on the validation set, and the performance is reported on the unseen testing set.
All the experiments in this paper are conducted on an NVIDIA A100 80GB GPU.

\noindent\textbf{Baseline Methods.} For a fair comparison, we take sampling-based methods as our baseline methods while excluding sketch-based ones, as our method accesses only a portion of the column data rather than the full table. We present the baselines with brief descriptions below, and the heuristics or assumptions of these methods will be discussed in Section~\ref{sec:relatedwork}.

\begin{itemize}[leftmargin=10pt]
    \item \texttt{Goodman}~\cite{goodman1949estimation} is the seminal NDV estimation method and it has the expression in Equation~(\ref{eq:goodman}).
    \item \texttt{GEE}~\cite{gee_charikar2000towards} estimates NDV by: $\hat{D}_{\mathrm{\texttt{GEE}}}=\sqrt{N/n}f_1+\sum_{j=2}^nf_j$.
    \item \texttt{EB}~\cite{error_bound} improves \texttt{GEE}: $\hat{D}_{\mathrm{\texttt{EB}}}=\sqrt{N/n}f_1^++\sum_{j=2}^nf_j,f_1^+=\max(1,f_1)$.
    \item \texttt{Chao}~\cite{chao1984nonparametric,chao_in_db_ozsoyoglu1991estimating} has a nonlinear polynomial expression: $\hat{D}_{\mathrm{\texttt{Chao}}}=d+f_1^2/2f_2$. If $f_2$ is zero, we will use $d$ as the estimation.
    \item \texttt{Shlosser}~\cite{shlosser1981estimation}: $\hat{D}_{\texttt{Shlosser}} = d + \frac{f_1 \sum_{i=1}^n (1 - r)^i f_i}{\sum_{i=1}^n ir(1 - r)^{i-1} f_i}$.
    \item \texttt{Jackknife}~\cite{burnham1978estimation,burnham1979robust}. Denote $d_{n-1}(k),1\leq k \leq n$, $d_{n-1}(k)=d_n-1$ if the attribute value for tuple $k$ is unique; otherwise $d_{n-1}(k)=d_n-1$. The first-order \texttt{Jackknife} method is: $\hat{D}_{\mathrm{\texttt{Jackknife}}}=d_n-(n-1)(d_{n-1}-d_n)$.
    \item \texttt{Sichel}~\cite{sichel1986parameter,sichel1986word,sichel1992anatomy} method needs to solve non-linear equations and it is shown in Equation~(\ref{eq:sichel}).
    \item \texttt{Bootstrap}~\cite{bootstrap_smith1984nonparametric}: $\hat{D}_{\texttt{Boot}}=d+\sum_{j:n_j>0}(1-n_j/n)^n$. It may perform worse when $D$ is large and $n$ is small because $\hat{D}_{\texttt{Boot}}\leq 2d$.
    \item \texttt{HT}~\cite{horvitz_sarndal1992model} defines $h_n(x)=\frac{\Gamma(N-x+1)\Gamma(N-n+1)}{\Gamma(N-n-x+1)\Gamma(N+1)}$, where $\Gamma$ is the gamma function, and $\hat{D}_{\texttt{HT}}=\sum\limits_{j:n_j>0}\frac{1}{1-h_n(\hat{N}_j)}$, $\hat{N}_j=N(n_j/n)$. 
    \item The family of \texttt{MoM}~\cite{mmo_bunge1993estimating}. \texttt{MoM1} needs to solve $d=\hat{D}_{\texttt{MoM1}}(1-e^{-n/\hat{D}_{\texttt{MoM1}}})$. \texttt{MoM2}: $d=\hat{D}_{\texttt{MoM2}}(1-h_n(N/\hat{D}_{\texttt{MoM2}}))$ where $h_n(\cdot)$ is defined identically to that in \texttt{HT}. 
    \item \texttt{LS}~\cite{ls_wu2022learning} is the first method based on ML techniques. It is pre-trained on a manually crafted dataset with 0.72 million data points in which the frequency profile of the columns follows specific distributions. For a fair and comprehensive comparison, we fine-tune \texttt{LS} on our training set, and it is denoted as \texttt{LS(FT)}.
\end{itemize}

Some baseline methods (\texttt{ChaoLee}~\cite{chaolee}, \texttt{MoM3}~\cite{mmo_bunge1993estimating}, and \texttt{SJ}~\cite{hybskew_haas1995sampling}), have extremely sophisticated expressions. Due to space limitations, we omit the equations in this paper and refer to~\cite{ndvlib,spang2019estimating} for the details. 

\noindent\textbf{Research Questions.} We conduct experiments to answer the following Research Questions (RQs).
\begin{itemize}[leftmargin=10pt]
    \item RQ1: What is the performance of baselines and the proposed \texttt{PLM4NDV} under sequential access and random sampling conditions when the sample sizes are the same? Can \texttt{PLM4NDV} significantly improve estimation accuracy under limited data scenarios? 
    \item RQ2: We argue that \texttt{PLM4NDV} largely relies on semantic features from the schema. Does the semantic information contained in the schema significantly contribute to NDV estimation? Besides, how useful is the semantic information from other columns in the table?
    \item RQ3: What is the effect of different fine-tuned PLMs in column representation for NDV estimation? How sensitive is \texttt{PLM4NDV} to hyperparameters?
    \item RQ4: We claimed that the distinct advantage of \texttt{PLM4NDV} is its applicablity even without data access. What is the performance in this scenario? How do \texttt{PLM4NDV} and the baselines perform with smaller sample sizes? Can \texttt{PLM4NDV} still outperform sampling-based baseline methods when randomly sampling a considerable data volume?
    \item RQ5: Efficiency significantly affects the practicality of NDV estimation methods. What are the efficiency and actual time consumption of \texttt{PLM4NDV}?
    \item RQ6: In the sequential access setting, the data layout of the original table may introduce bias. How does data layout impact NDV estimation under sequential access scenarios?
\end{itemize}

\begin{table*}[t]
    \centering
    \caption{Mean and quantiles of q-error performance on the test set under sequential access first 100 rows per column. The best-performing metrics are highlighted in boldface.}
    \begin{tabular}{c|cccccc}
\toprule

    Method & Mean & 50\% & 75\%& 90\% & 95\% & 99\%   \\
\midrule
    \texttt{Goodman} & $\infty$ &  3.46  & 73.50  & 1.32e3 &  7.56e15 & $\infty$  \\
    \texttt{GEE} & 125.49 & 4.00 & 14.54 & 49.90 & 154.60 & 1.60e3 \\
    \texttt{EB} & 17.70 & 7.38 & 15.14 & 30.07 & 51.09 & 186.41 \\
    \texttt{Chao} & 225.09 & 14.20 & 150.00  &  474.21 &  964.15 & 3.24e3  \\
    \texttt{Shlosser} & 129.23 & {2.04}  & 10.17 & 61.00  & 191.00  & 1.71e3  \\
    \texttt{Jackknife} & 205.39 & 7.07 & 76.53 & 265.84 &  603.82 & 3.02e3 \\
    \texttt{Sichel} & 249.14 & 6.34 & 109.10 & 389.66 & 822.36 &  3.20e3  \\
    \texttt{Bootstrap} & 86.63 & 27.77 & 70.34 & 143.71 & 300.08 & 1.09e3 \\
    \texttt{HT} & 1.73e3 & 27.22 & 381.99 & 2.93e3 & 7.08e3 & 2.96e4 \\
    \texttt{MoM1} & 3.01e5 & 6.20  & 104.00  & 1.67e5  &  4.39e5  & 3.38e6 \\
    \texttt{MoM2} & 163.69 & 6.50 & 26.32 & 117.59 & 351.00 & 2.45e3 \\
    \texttt{MoM3} & 169.98 & 7.93 & 37.27 & 148.53 & 360.23 & 2.45e3 \\
    \texttt{ChaoLee} & 278.88 & 9.64 & 128.67 & 445.40 & 993.71 & 3.49e3 \\
    \texttt{SJ} & 159.33 & 2.20 & 16.67 & 118.81 & 350.97 & 2.39e3 \\
    \texttt{LS} &  43.05 &  2.90  &  6.87  &  31.08  &  93.35  &  492.48 \\
    \texttt{LS(FT)} &  22.39 &  3.30  &  5.96  &  17.32  &  48.70  &  321.01 \\
\midrule
    \texttt{PLM4NDV} & \textbf{13.33} &  \textbf{1.86} & \textbf{3.81}  & \textbf{10.81}  &  \textbf{25.18} &  \textbf{148.76}  \\
\bottomrule
    \end{tabular}
    \label{tab:mainexp}
\end{table*}

\begin{table*}[t]
    \centering
    \caption{Mean and quantiles of q-error performance on the test set under random sampling 100 rows per column. The best-performing metrics are highlighted in boldface.}
    \begin{tabular}{c|cccccc}
\toprule

    Method & Mean & 50\% & 75\%& 90\% & 95\% & 99\% \\
\midrule
    \texttt{Goodman}  & $\infty$  &  1.79  &  28.78  &  680.55 & 3.92e17 & $\infty$ \\
    \texttt{GEE} & 7.06 & 2.74  &  7.94  &  16.09  &  22.90  &  51.60\\
    \texttt{EB}  &  9.54  &  6.15  &  12.15  &  20.59  &  29.28  &  52.78\\
    \texttt{Chao}  & 210.76 &  3.27 &  123.22 &  449.78 &  962.46 &  3.27e3\\
    \texttt{Shlosser}  & 22.35 &  2.00  &  10.41  &  43.07  &  87.30  & 335.26 \\
    \texttt{Jackknife} & 72.57 &  2.25  &  41.16  &  143.26  &  303.14  &  1.16e3\\
    \texttt{Sichel} & 141.49 &  2.16  &  88.58  &  300.31  & 613.23 &  2.16e3 \\
    \texttt{Bootstrap} & 83.52 &  25.00  &  63.94  &  137.93  &  290.58  &  1.09e3\\
    \texttt{HT}  &  1.35e3 &  20.45  &  408.79  &  2.99e3  &  6.73e3  &  2.28e4\\
    \texttt{MoM1} & 4.90e4 &  2.00 &  12.30 &   1.49e5 &   3.73e5 &  6.78e5\\
    \texttt{MoM2} & 18.81 &  2.00  &  12.38  &  28.32  &  46.70  &  223.77\\
    \texttt{MoM3} & 26.50&  2.54 &  18.02 &  57.10 &  107.62 & 242.63\\
    \texttt{ChaoLee} & 141.61 & 3.30  & 76.19  & 280.18  & 598.08  & 2.32e3\\
    \texttt{SJ} & 11.95 &  1.31  &  2.99  &  11.29  &  26.63  &  171.82\\
    \texttt{LS} & 3.74 & 2.62 &  4.09  &  6.48  &  9.27  &  16.77\\
    \texttt{LS(FT)}  & 2.69 &  1.49  &  2.06  &  3.55  &  5.09  &  14.27 \\
    \texttt{PLM4NDV} & \textbf{2.17} &  \textbf{1.30}  &  \textbf{1.76 } &  \textbf{2.84}  &  \textbf{4.26}  &  \textbf{11.64}   \\
\bottomrule
    \end{tabular}
    \label{tab:mainexp2}
\end{table*}

\subsection{Main Results (RQ1)}\label{sec:exp:rq1}
For the same sample size, we evaluate the performance of \texttt{PLM4NDV} and baselines using both the first 100 rows accessed sequentially and 100 rows sampled randomly from each column, respectively. To comprehensively illustrate the performance of the methods, we present both the mean and various percentiles of q-error on the test set. The results are shown in Table~\ref{tab:mainexp} and Table~\ref{tab:mainexp2}, where $\infty$ indicates that the number exceeds the representation limits of a 32-bit floating-point type. Based on the results, we can draw the following conclusions:

\begin{itemize}[leftmargin=15pt]
    \item \texttt{PLM4NDV} achieves the best performance across all metrics under both sequential access and random sampling conditions. In sequential access, the majority (90\%) of test cases have a q-error below 10.81, while in the random sampling setting, this metric is 2.84, which is significantly lower than that of other methods under both conditions. The consistently superior performance across all metrics and substantially low q-error in the majority of test cases demonstrate the effectiveness of our method.
    \item Under the sequential access condition, estimating NDV under limited data is challenging for sampling-based baseline methods. On the one hand, we observe that most of these methods have substantial NDV estimation q-errors. On the other hand, the growth between the quantiles of q-error is quite rapid, indicating unstable performance under limited data scenarios. Although a few methods (\texttt{Shlosser},\texttt{SJ}, and \texttt{LS}) exhibit promising results in the 50\% quantile of q-error that is less than 3, their poor performance in certain scenarios significantly diminishes their overall effectiveness.
    \item Under the random sampling condition, nearly all methods show a significant reduction in estimation error across each metric, indicating the importance of IID samples, as previously noted in prior approaches. This is because sequentially accessed samples may not be IID and can be influenced by the data layout, so we will investigate the impact of the data layout in Section~\ref{sec:exp:layout}. The overall performance of \texttt{PLM4NDV} is still the best among all methods, indicating that incorporating semantic information is also effective under the IID samples condition.
    \item The number of training data points for the learning-based baseline methods (\texttt{LS} and \texttt{LS(FT)}) is approximately twenty times greater than that of \texttt{PLM4NDV}. The fine-tuned version exhibits performance improvement over \texttt{LS} in most metrics (except for the 50\% percentile of q-error under sequential access scenarios), indicating the difficulties of NDV estimation under sequential data access scenarios. \texttt{LS(FT)} exhibits inferior performance than \texttt{PLM4NDV}, demonstrating the significance of leveraging semantics in NDV estimation.
    \item Notably, \texttt{PLM4NDV}, utilizing sequentially accessed data, outperforms most baselines even under random sampling conditions across most metrics. This highlights that when the sampling cost budget is limited, \texttt{PLM4NDV} using sequential access is practical enough because its performance is comparable to most baselines using random sampling.
    \item In summary, our method \texttt{PLM4NDV} consistently outperforms existing approaches with the same data access costs. Additionally, \texttt{PLM4NDV} achieves higher estimation accuracy with lower data access costs. The superiority of \texttt{PLM4NDV} demonstrates the significant benefit of incorporating semantic information for NDV estimation.
\end{itemize}

\begin{figure}
    \centering
    \includegraphics[width=\linewidth]{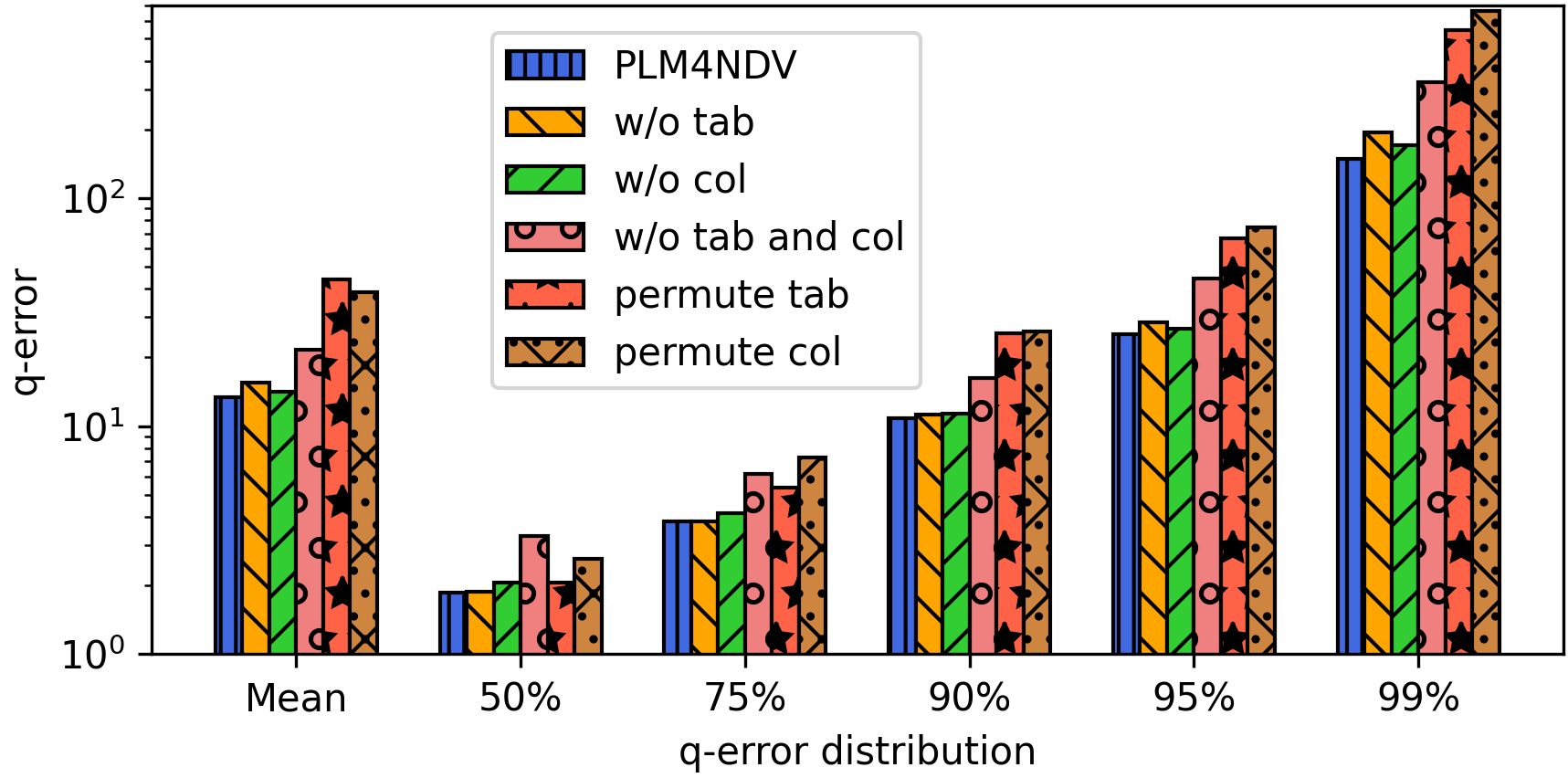}
    \caption{Ablation study on semantic features evaluated under the sequential access condition, where ``w/o'' indicates removing a kind of semantic features from \texttt{PLM4NDV}.}
    \label{fig:semantic}
    \vspace{-3mm}
\end{figure}

\subsection{Effect of Semantic Features (RQ2)}\label{sec:rq2semantic}
To verify the effectiveness of semantic features, we test the following variants of \texttt{PLM4NDV}. The variant ``w/o col'' indicates removing the column embedding $x$. Similarly, removing the interacted column information $x^\prime$ is denoted as ``w/o tab''. Dropping all semantic features is represented as ``w/o tab and col''. These three variants of \texttt{PLM4NDV} thoroughly explore the effect of semantic features from the column itself and the auxiliary of the table. In addition, we construct two variants to distort the semantics: ``permute col'' refers to the random permutation of the textual tokens in the serialized schema, intending to disrupt the semantics of each column. ``permute tab'' denotes the random permutation of column texts within the same table, which undermines the relationship between column semantics and column data.
Their performance evaluated under the sequential access condition is depicted in Figure~\ref{fig:semantic}, from which we can draw the following conclusions:
\begin{itemize}[leftmargin=15pt]
    \item Semantic information significantly contributes to NDV estimation. The variant ``w/o tab and col'' solely leverages the frequency profile from the sampled data to establish the estimation. It is obvious that removing or distorting the semantic features of both the column and the table leads to a significant increase in the mean and each quantile of q-error. This illustrates the effectiveness of using semantic features from the schema in NDV estimation.
    \item The auxiliary semantics of the columns within the same table are critical. Removing the auxiliary semantics from the table, the variant ``w/o tab'' relies on the column-independent semantic information in the estimation. We observe a consistent decline in the estimation performance, highlighting the significance of the table schema in NDV estimation and the effectiveness of the table semantics component in \texttt{PLM4NDV}. 
    \item The semantics of the whole table are useful for estimations. The variant ``w/o col'' performs better than ``w/o tab and col'' but is comparable to ``w/o tab''. On the one hand, this indicates that the semantics within the table, combined with the object column's profile, can provide some useful context, as the semantics of the target column are inherently contained within the semantics of the table to some extent. On the other hand, relying solely on the table information does not fully convey the meaning of the object column, leading to a decline in performance compared to \texttt{PLM4NDV}. This further emphasizes the importance of understanding the semantic meanings of the object columns in NDV estimation.
    \item Using incorrect semantics (``permute col'' and ``permute tab'') significantly increases the mean, 90\%, 95\%, and 99\% q-error compared to scenarios without semantics (``w/o tab and col''). This indicates that the accuracy of semantics is crucial for \texttt{PLM4NDV}, and its superior performance relies on effectively leveraging semantics. Among the two variants using incorrect semantics, ``permute col'' performs worse than ``permute tab'' in the percentiles of q-error. This difference can be explained as follows: a random permutation of the textual tokens in the serialized column schema may destroy the semantics of both columns and tables. For instance, a column with the serialized text ``EmployeeID,int'' could be transformed into ``Il,etpeyDiomnE'' which is semantically meaningless. On the contrary, the mean q-error of ``permute tab'' is greater than ``permute col'', indicating that using the semantics of other columns may bring significant errors in extreme scenarios that increase the mean q-error. For instance, a column stores the data of unique identifiers but is named with the semantics of gender.
    \item The above findings reveal that the effectiveness of our method stems from the combination of column and table semantic information. Relying on only one of them or using incorrect semantics does not yield optimal estimation results.
\end{itemize}

\subsection{Robustness Analysis (RQ3)}

\begin{figure}
    \centering
    \includegraphics[width=\linewidth]{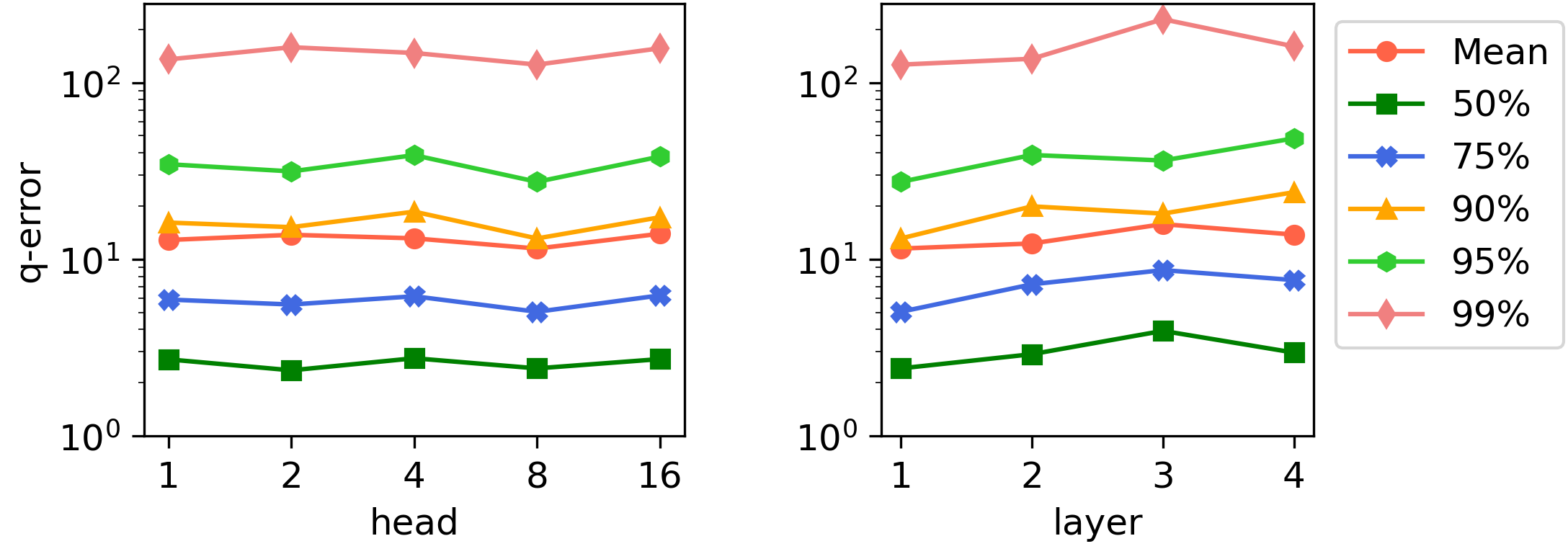}
    \caption{Hyperparameter study of \texttt{PLM4NDV}.}
    \label{fig:hyper}
\end{figure}

\noindent\textbf{Impact of hyperparameters.} There are two hyperparameters in \texttt{PLM4NDV}: the number of heads and layers in the MHSA column interaction component. We search the number of attention heads in $\{1,2,4,8,16\}$ and the number of layers in $\{1,2,3,4\}$ to investigate the impact of them, the performance of \texttt{PLM4NDV} on the test set under different settings is depicted in Figure~\ref{fig:hyper}. Based on the results in the figure, we can derive the following findings.

\begin{itemize}[leftmargin=15pt]
    \item The overall performance of \texttt{PLM4NDV} is not highly sensitive to the number of heads, with variations in the number of heads not resulting in significant performance differences. \texttt{PLM4NDV} demonstrates the optimal overall performance when the number of heads is 8.
    \item Increasing the number of layers leads to a decline in estimation accuracy, with the optimal estimation performance across all metrics obtained when the number of layers is 1. A possible reason is that the latent semantic information between columns can be captured effectively with just one interaction layer, and adding more layers may lead to overfitting.
    \item Although adjusting the hyperparameters results in variations in performance, we can conclude that \texttt{PLM4NDV} is not sensitive to the two hyperparameters because the overall performance of \texttt{PLM4NDV} under different settings outperforms most baseline methods. Therefore, the number of heads and layers is set to 8 and 1, respectively.
\end{itemize}

\begin{figure}
    \centering
    \includegraphics[width=\linewidth]{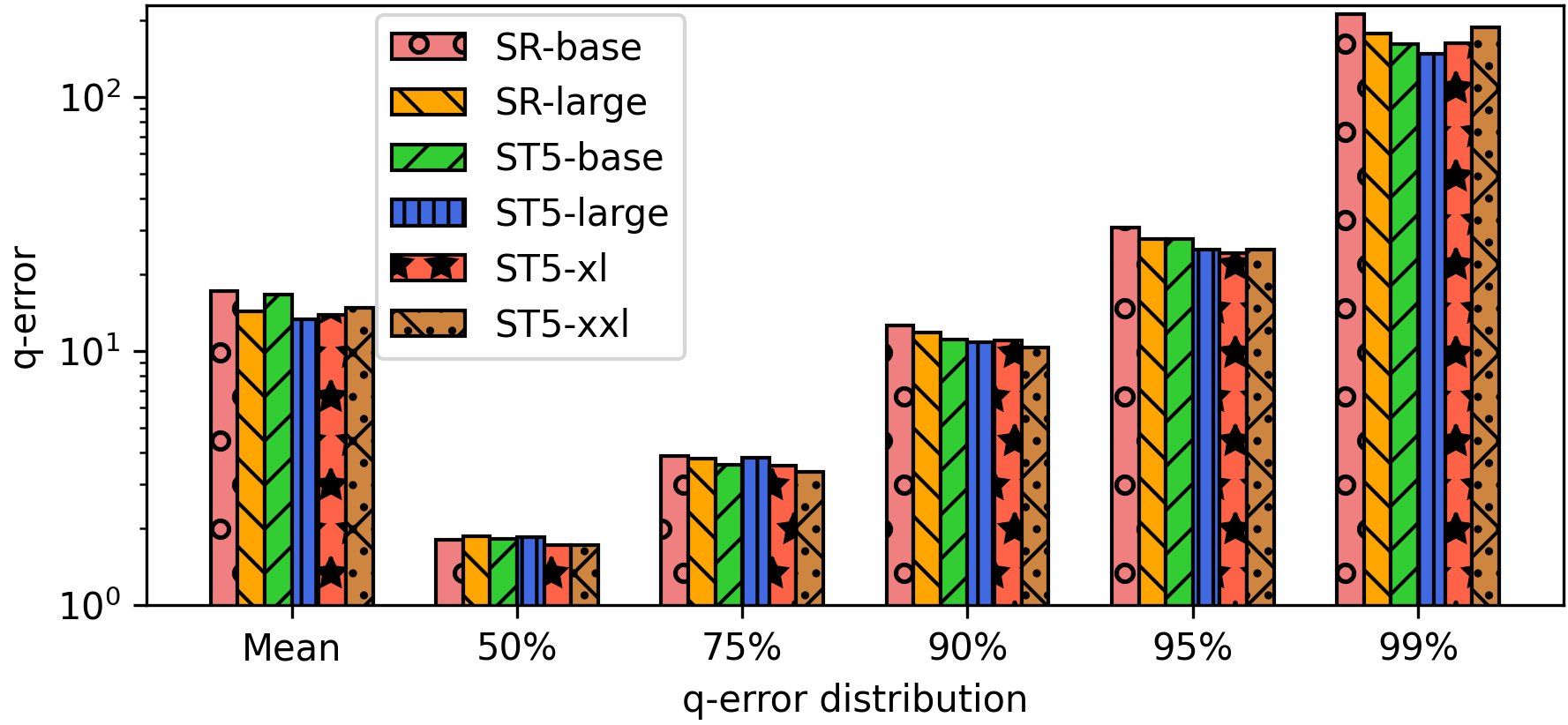}
    \caption{Performance comparison of six PLMs as the foundation PLM in \texttt{PLM4NDV}.}
    \label{fig:embedding}
\end{figure}

\noindent\textbf{Impact of foundation PLMs.} Due to the numerous foundational PLMs available, we selected two representative PLMs that are well fine-tuned for the task of semantic textual similarity: Sentence-RoBERTa (\texttt{SR})~\cite{reimers2019sentence} and Sentence-T5 (\texttt{ST5})~\cite{ni2021sentencet5}. These are fine-tuned versions of RoBERTa~\cite{roberta} and T5~\cite{t5_raffel2020exploring}, respectively. We chose these two fine-tuned PLMs due to their effectiveness and generalization capabilities in encoding serialized schema sentences.
We study the impact of PLMs with different sizes, and the number of parameters of the PLMs discussed in this section are shown in Table~\ref{tab:size}. The performance of using different PLMs as the column embedding component in \texttt{PLM4NDV} under sequential access scenario is depicted in Figure~\ref{fig:embedding}. Based on the sizes and performance of different PLMs, we can draw the following conclusions:

\begin{table}[t]
    \centering
    \caption{Number of parameters of the discussed foundation PLMs, where `M' and `B' represent million and billion, respectively.}
    \begin{tabular}{ccccccc}
\toprule
    {SR-base} & {SR-large} & {ST5-base} & {ST5-large} & {ST5-xl} & {ST5-xxl} \\
\midrule
    125M   & 355M   & 110M & 335M &  1.24B & 4.86B \\
\bottomrule
    \end{tabular}
    \label{tab:size}
\end{table}

\begin{itemize}[leftmargin=15pt]
    \item The exploitation of the column embeddings obtained from different foundation PLMs yields varying estimation results in \texttt{PLM4NDV}, with significant differences observed in the log-scaled q-errors depicted in the figure. This suggests that the semantic information extracted by different PLMs directly impacts NDV estimation and underscores the importance of selecting an appropriate PLM in our method for extracting semantic information from the schema.
    \item For the \texttt{SR} models, using \texttt{SR-large} leads to an improvement in estimation performance than that of \texttt{SR-base} in most metrics. The two variants of \texttt{SR} models have different sizes but similar architectures, \texttt{SR-large} exhibits better performance because it has a larger number of parameters, allowing for more effective semantic representation compared to \texttt{SR-base}, thereby improves the estimation accuracy.
    \item For the \texttt{ST5} models, in the majority of test cases (the 90\% percentile), increasing the model parameters improves estimation accuracy, except for \texttt{ST5-xl}. Using \texttt{ST5-xxl} as the semantic embedding model has the optimal performance in the 50\%, 75\%, 90\%, and 95\% percentile of q-error, while the worse performance in a few cases increases its mean q-error. In general, using \texttt{ST5-xxl} has the optimal performance in most metrics. This shares the same reason as \texttt{SR-large} that a larger model size leads to more effective semantic representations. 
    \item Compared to the \texttt{SR} and \texttt{ST5} models, we can observe that the \texttt{ST5} models exhibit better performance than \texttt{SR} models where the number of parameters is similar. There are two possible reasons for the situation. On the one hand, the two PLMs are fine-tuned in different policies and datasets, resulting in different semantic embedding capabilities. On the other hand, T5 has an Encoder-Decoder architecture in the pre-training stage, which may make its encoder (\texttt{ST5} models) have better semantic representation abilities compared to RoBERTa models, which are trained in the Encoder-only architecture. 
    \item The above analysis demonstrates that the column embeddings obtained from different PLMs significantly impact NDV estimation. Nonetheless, even the lowest performance among the six variants still outperforms all baseline methods in the majority of test cases. Considering both estimation accuracy and model efficiency, we use \texttt{ST5-large} as the foundation PLM in \texttt{PLM4NDV} rather than the best-performing \texttt{ST5-xxl}. This analysis also highlights that effectively fine-tuning PLMs in the context of database schemas may be a promising area for future exploration.
    \item In addition, the T5-11B version is regarded as a pioneering Large Language Model (LLM) work in the NLP community~\cite{zhao2023survey}, with \texttt{ST5-xxl} leveraging the encoder component of the T5-11B version. Since the above analysis reveals that the PLMs with more parameters may have better semantic representation abilities that bring better NDV estimation performance, we will investigate and discuss the effect of prompting LLMs with larger sizes for NDV estimation in Section~\ref{sec:exp:llm}. 
\end{itemize}

\subsection{Access Data Volume Discussion (RQ4)}

\begin{table}[t]
    \centering
    \caption{The 90\% quantile of q-error of each method under different data volumes, where ``N/A'' represents not applicable, and $\downarrow$ indicates the performance is inferior to \texttt{PLM4NDV} without data access (underlined).}
    \begin{tabular}{ccccc|c}
\toprule
    & n=0 &  n=10 & n=20 & n=50 & n=0.01N \\
\midrule
    \texttt{Goodman} & N/A & 1.32e4$\downarrow$   & 5.53e4$\downarrow$ & 3.20e4$\downarrow$  & 101.15$\downarrow$\\
    \texttt{GEE}  & N/A & 49.90  & 88.14$\downarrow$ & 63.00$\downarrow$ &  10.00 \\
    \texttt{EB}  & N/A & 30.07  & 57.31 & 39.44 & 10.14 \\
    \texttt{Chao} & N/A & 474.21$\downarrow$  & 1.64e3$\downarrow$ & 775.82$\downarrow$ & 100.17$\downarrow$\\
    \texttt{Shlosser}  & N/A & 61.00$\downarrow$   & 264.28$\downarrow$ & 104.00$\downarrow$ &  8.11\\
    \texttt{Jackknife}  & N/A & 265.84$\downarrow$  & 1.10e4$\downarrow$ & 500.60$\downarrow$ & 50.06 \\
    \texttt{Sichel}  & N/A & 389.66$\downarrow$  & 1.62e3$\downarrow$ & 720.02$\downarrow$ &  100.65$\downarrow$\\
    \texttt{Bootstrap}  & N/A & 143.71$\downarrow$  & 731.00$\downarrow$ & 288.57$\downarrow$ &  243.50$\downarrow$ \\
    \texttt{HT} & N/A & 2.93e4$\downarrow$  & 3.24e4$\downarrow$ & 2.92e4$\downarrow$ & 6.33e3$\downarrow$ \\
    \texttt{MoM1}  & N/A & 1.67e5$\downarrow$  & 5.26e5$\downarrow$ & 1.37e5$\downarrow$ & 11.88 \\
    \texttt{MoM2}  & N/A &  117.59$\downarrow$  & 252.94$\downarrow$ & 184.46$\downarrow$ & 13.72 \\
    \texttt{MoM3}  & N/A & 148.53$\downarrow$ & 253.58$\downarrow$ & 188.41$\downarrow$ & 1.18e3$\downarrow$ \\
    \texttt{ChaoLee}  & N/A & 445.40$\downarrow$  & 1.92e3$\downarrow$ & 848.43$\downarrow$ & 98.78$\downarrow$ \\
    \texttt{SJ}  & N/A &  118.81$\downarrow$ & 366.0$\downarrow$ & 191.37$\downarrow$ & 32.41 \\
    \texttt{LS}  & N/A & 31.08  & 33.73 & 24.19 & 6.48 \\
    \texttt{LS(FT)}  & N/A & 43.49  & 33.09 & 22.77 & 2.04 \\
\midrule
    \texttt{PLM4NDV} & \underline{\textbf{59.51}} & \textbf{25.74} & \textbf{20.67} & \textbf{15.25} &  \textbf{1.89} \\
\bottomrule
    \end{tabular}
    \label{tab:volume}
\end{table}

We compare the performance of each method under different data access volumes and discuss the 90th percentile of q-error to demonstrate their performance in the majority of the test cases, as shown in Table~\ref{tab:volume}. Based on the results presented in the table, we proceed with the following discussions:

\noindent\textbf{Estimation without data access.} There are scenarios where estimation must be performed without sampling data, particularly in cases where accessing databases is prohibited or the data access cost budget is insufficient even for a single sample. Existing methods are not applicable when data is inaccessible, conversely, \texttt{PLM4NDV} is uniquely suited for this scenario as it can solely utilize the semantic features for estimation. The 90th percentile of the q-error for \texttt{PLM4NDV} when estimating NDV without data is 59.51. While this indicates a relatively high estimation error under this scenario, the performance of \texttt{PLM4NDV} is significant from two perspectives. First, in the scenario of estimation without data access, our method is the only one that functions. Second, by comparing the results with those obtained from accessing 100 rows in Table~\ref{tab:mainexp}, it is observed that \texttt{PLM4NDV} without accessing any data can even outperform most methods when accessing 100 rows, whether in sequential or random order. Therefore, the estimation performance of \texttt{PLM4NDV} in the no-data-access scenario is promising.

\noindent\textbf{Varying sequential access volumes.} We study the performance of baselines and our method for $n=\{10,20,50\}$ to investigate the effect of different sequential access volumes under limited data scenarios. Referring to the results in Table~\ref{tab:mainexp} and Table~\ref{tab:volume}, we can make the following observations. For all baseline models, increasing the data volume does not necessarily lead to performance improvements. Besides, \texttt{LS(FT)} shows a performance decline compared to \texttt{LS} when $n$ is 10 and 20, demonstrating the challenges in estimating under limited data access. On the contrary, the estimation error of \texttt{PLM4NDV} consistently declines as the accessed volume increases, and the best performance in each scenario remains with \texttt{PLM4NDV}. 

\noindent\textbf{Random sampling 1\%.} To further examine whether the conclusions regarding \texttt{PLM4NDV} remain valid when random sampling a considerable amount of data, we explore the performance in the scenario of sampling 1\% of the full data uniformly at random, which is typically required by sampling-based methods. Given that the frequency profile varies with different data lengths, we use a cut-off of the first 100 frequency profiles in \texttt{PLM4NDV} for this scenario. Referring to the results in Table~\ref{tab:mainexp} and Table~\ref{tab:volume}, we can observe that most methods show substantial improvement compared to scenarios with limited data. However, despite randomly sampling a considerable amount of data, the performance of some methods remains inadequate, indicating the need for an even larger sample size. In contrast, \texttt{PLM4NDV} exhibits a significant performance improvement when accessing a considerable amount of data, with a 90\% percentile of q-error of only 1.89, indicating that its estimation error can be substantially reduced in the majority of the test cases. This outcome highlights the practicality of incorporating semantic information in \texttt{PLM4NDV}. In addition, the performance of \texttt{PLM4NDV} without data access even beats many methods that randomly sample 1\% of the full data, highlighting the distinct promising advantages of \texttt{PLM4NDV}.

The results in Table~\ref{tab:mainexp}, Table~\ref{tab:mainexp2}, and Table~\ref{tab:volume} demonstrate that utilizing semantics can enhance NDV estimation in both sequential access and random sampling scenarios.

\subsection{Efficiency Analysis (RQ5)}\label{sec:exp:efficiency}

The efficiency of \texttt{PLM4NDV} encompasses two phases: training and inference. Most baseline sampling-based methods are purely statistical and can be applied directly, with only \texttt{LS} involving a training stage. Since \texttt{PLM4NDV} is the only one that utilizes semantic information, there is no direct basis for comparison with other methods in the training phase. Therefore, we intuitively present the time consumption during training. For the inference stage, we demonstrate the average time consumption in each procedure of \texttt{PLM4NDV}, and the results are shown in Table~\ref{tab:efficiency}. We record the time consumption on a common virtual machine with 8 cores and 16 GB of memory.

\noindent\textbf{Training.} The cost of constructing the frequency profiles is trivial, the majority of time consumption is using PLMs to obtain the column embeddings and training the learned NDV estimation model. Specifically, the column embedding time is approximately 30s and the training time is about 3100s. The time consumption for column embedding is minimal compared to model training, indicating that utilizing semantic information does not create an efficiency bottleneck. Furthermore, since \texttt{PLM4NDV} captures the semantic features of individual columns and their corresponding table using an attention model, the training process is relatively time-consuming. Nevertheless, the training is conducted offline and updated infrequently, making the cost acceptable in practice.

\begin{table}[t]
\caption{Average time consumption for an estimation on a column in each procedure of \texttt{PLM4NDV} in the inference stage.} \label{tab:efficiency}

\begin{tabular}{cccccc}
\toprule
       data access &  statistics & semantics & estimation \\
\hline
        $\sim$10ms & 1ms<  & \makecell{$\sim$2ms(GPUs) \\ $\sim$1.23s (CPUs) } & 1ms<\\
\bottomrule
\end{tabular}
\vspace{-3mm}
\end{table}

\noindent\textbf{Inference.} As shown in Table~\ref{tab:efficiency}, \texttt{PLM4NDV} significantly reduces data access costs, requiring only 10ms per access because it sequentially accesses a fixed number of samples rather than randomly sampling. Randomly sampling data from databases is extremely time-consuming~\cite{olken1993random}, and this cost increases with the size of the table. In contrast, the data access time of \texttt{PLM4NDV} is agnostic to the table size, consistently requiring about 10ms for each access.

The length of the frequency profiles of \texttt{PLM4NDV} is unaffected by the table size, allowing for high efficiency in statistics construction. It takes less than 1ms for a column.

\texttt{PLM4NDV} model involves two parts: semantic embedding and the learnable component estimation. It takes about 1.75s to load the PLM from disk for the first time. The average PLM embedding time is approximately 2ms on GPUs and about 1.23s on CPUs for a column. The average inference time for a column of the learnable model is less than 1ms on both GPUs and CPUs.

\noindent\textbf{Efficiency of baselines.} The time consumption of baselines in the estimation stage is minimal, as each method takes only a few milliseconds to estimate the NDV of a column using the frequency profiles as input. Although the baselines appear to have higher efficiency in the estimation phase compared to \texttt{PLM4NDV}, their efficiency in random sampling is extremely low, leading to overall efficiency that is inferior to our approach.

\noindent\textbf{Practicality analysis.} In practice, the model is trained offline, meaning that inference efficiency is the key factor determining its practicality. Additionally, the column semantic embeddings can be computed independently before estimating NDV, as the number of schemas within a database is typically limited, and the storage requirements of column embeddings are finite. Consequently, the computational cost is acceptable (even when using CPUs), as each column needs to be computed only once for storage and can be pre-computed during idle time, allowing for direct querying of column embeddings from the storage during NDV estimation. Ultimately, the time consumption of \texttt{PLM4NDV} in online NDV estimation is minimal, with the average end-to-end estimation for a column taking only tens of milliseconds on CPUs. Therefore, even with the use of PLMs to capture semantic information in the schema, \texttt{PLM4NDV} maintains promising practicality.

\noindent\textbf{Limitations of \texttt{PLM4NDV}.} \texttt{PLM4NDV} may be impractical for extremely wide tables due to memory limitations, as the space complexity of the column interaction component is $O(t^2)$, where $t$ is the number of columns of a table. However, the occurrence of such scenarios is infrequent in practical applications. Besides, \texttt{PLM4NDV} may struggle to meet the real-time NDV estimation requirements when it needs to compute semantic embeddings without GPUs.

\subsection{Study of Data Layout (RQ6)}\label{sec:exp:layout}
\begin{figure}[t]
    \centering
    \includegraphics[width=0.8\linewidth]{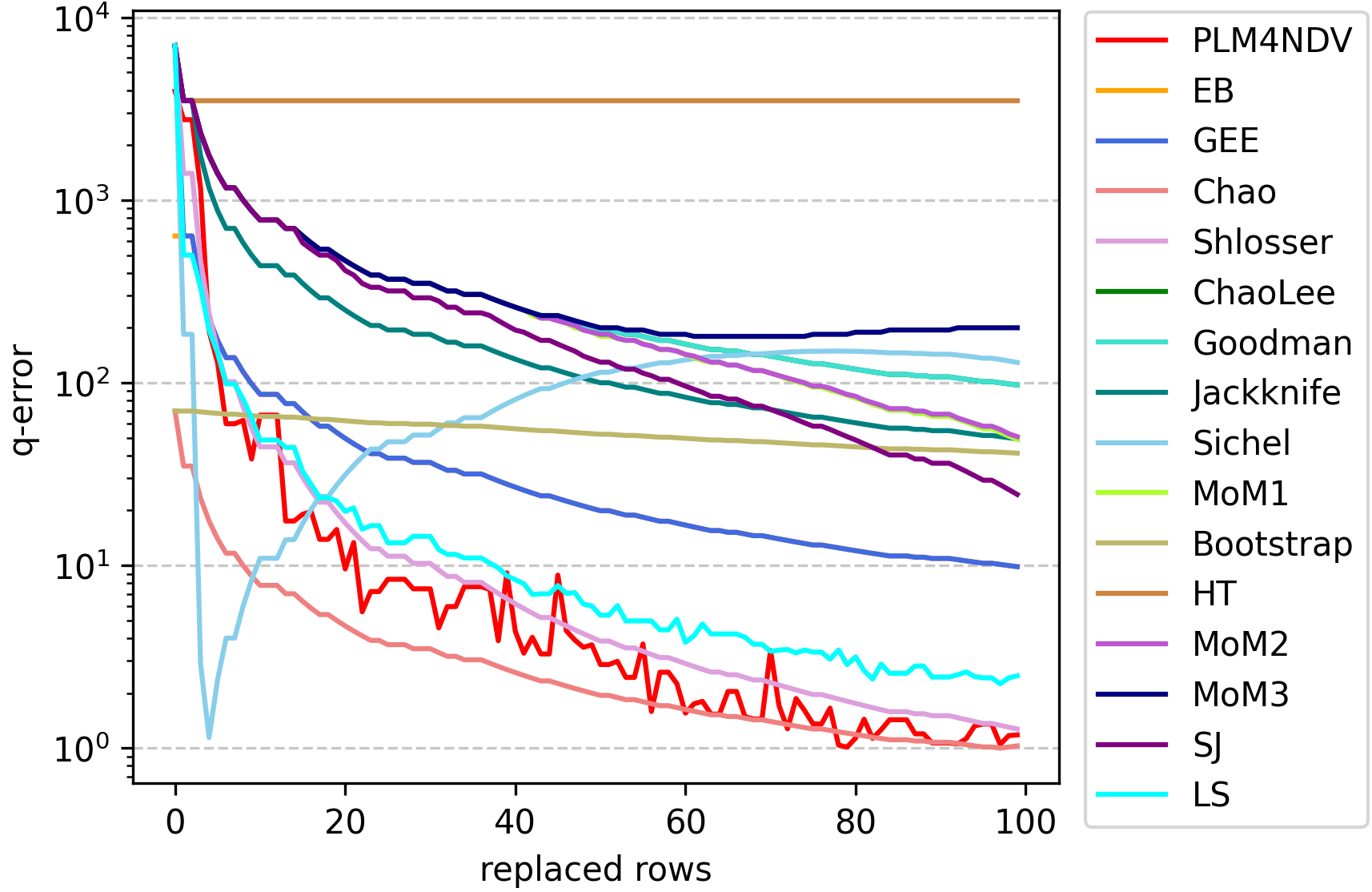}
    \caption{Estimation error variation with the number of replaced rows in the first 100 rows of the synthesis table.}
    \label{fig:layout}
    \vspace{-3mm}
\end{figure}

We construct a synthesis table with two columns: \lstinline[style=SQLStyle]{EmployeeID} and \lstinline[style=SQLStyle]{Office Address}, where the data types are \lstinline[style=SQLStyle]{int} and \lstinline[style=SQLStyle]{string}, respectively. We investigate the impact of the data layout of \lstinline[style=SQLStyle]{Office Address} column on NDV estimation. Denote $p=\frac{D}{N}$ is the selectivity of the column. Assuming that $N=10,000$ and we set a relatively high selectivity $p=0.7$ such that sequentially accessing the first 100 rows can introduce bias in NDV estimation. In this scenario, we consider that the first 3,001 rows share the same office address.

To explore the impact of data layout on the sequential access scenario, we gradually replace the first 100 rows one by one by randomly selecting a row from rows 101 to 10,000.  \texttt{PLM4NDV} is trained on the training set as described in Section~\ref{sec:setting}. The performance of baselines and \texttt{PLM4NDV} is shown in Figure~\ref{fig:layout} and we can derive the following conclusions: 
\begin{itemize}[leftmargin=15pt]
    \item The data layout of the table significantly affects the performance of most NDV estimation methods. For the synthesis column with a selectivity of 0.7, when the office addresses of the first 100 rows are identical (i.e., the number of replaced rows is 0), it can lead to significant estimation errors in most methods. 
    \item When increasing the randomness of the layout in the first 100 rows (i.e, increasing the number of replaced rows), some methods show an estimation error decrease, while a few methods remain unchanged. Besides, some methods may initially decrease and then increase, indicating that some baselines are more sensitive to data layout. The proposed method \texttt{PLM4NDV} tends to achieve a substantial reduction in estimation error and gradually outperforms most baselines. 
    \item Estimating NDV under sequential access the top 100 rows is significantly influenced by data layout. However, based on the experimental results presented in Table~\ref{tab:mainexp}, when the sampling budget allows accessing limited data, \texttt{PLM4NDV} emerges as the most effective alternative among existing methods.
\end{itemize}

\begin{lstlisting}[style=prompt, label=lst:method:column, caption={Template of prompting LLMs for NDV estimation. Some rules are omitted for space limitations. }, float=t]
You are a data expert, your task is to estimate the number of distinct values of a column by the following rules:
(1) The number of distinct values of the column must be smaller than or equal to the total rows. 
...

Here are the target column information:
Column name: [ColumnName].  Column type: [ColumnType].
Constraints: [ColumnConstraints]. Comment: [ColumnComment]. Total rows: [ColumnSize].
Data samples: [Samples]
Information of the table: [TableInfo]
<Output Format>
{
"Reason": "XXX",
"NDV": "XXX"
}
<\Output Format>
Try your best to guess the number of distinct values of the column. In any case, give an exact number of NDV. Let's think step by step in order to produce the answer. The number of distinct values of the column is:
\end{lstlisting}

\section{Preliminary Exploration of LLMs}
Large language models (LLMs) such as GPT4~\cite{achiam2023gpt4} and Llama~\cite{touvron2023llama} have achieved notable success 
 in NLP, demonstrating impressive zero-shot reasoning abilities across various domains and scenarios owing to their robust generalization and adaptability. 
These models possess significantly larger parameters than earlier PLMs like RoBERTa~\cite{roberta} and T5~\cite{t5_raffel2020exploring}, which were studied in this paper. Given these advancements, it is natural to inquire whether LLMs can be leveraged to provide better semantic understandings and improve NDV estimation accuracy. 

Fully leveraging LLMs in databases involves costly prompt engineering or even fine-tuning~\cite{gao2023text,gu2023few,gu2023interleaving,pourreza2024din,singh2023format5,huang2024llmtune,zhou2023llm,zhou2024dbot,zhou2024llm}, which is not the focus of this paper. In this section, we conduct a preliminary exploration of LLMs to validate the significance of utilizing semantics in NDV estimation and show some insightful findings for future works.

\subsection{Prompting LLMs for NDV Estimation}
One of the most fascinating aspects of LLMs is their ability to move beyond the fine-tuning paradigm~\cite{ding2022delta} that was often required by PLMs to adapt to downstream tasks. Instead, LLMs have shifted towards the prompt-tuning paradigm~\cite{liu2023pre}. There are multiple categories of methods in the prompt-tuning paradigm, we leverage the widely used \textit{tuning-free prompting} paradigm (\textit{\textbf{prompting}} for brevity) to adapt LLMs in the task of NDV estimation. Given that NDV estimation is not a typical NLP task, we formulate NDV estimation as a natural language question by constructing a simple prompt template, and the LLMs are requested to answer the question based on the provided schema and sampled data records.

\noindent\textbf{Prompt template.} The prompt template construct in this paper is illustrated in Listing~\ref{lst:method:column}, where the slots are enclosed in square brackets, and each slot is designed to be filled with specific variables. The column information corresponds to the schema and \lstinline[style=prompt]{ColumnSize} represents the size of the column. The \lstinline[style=prompt]{Samples} and \lstinline[style=prompt]{TableInfo} represent the accessed samples in the column and the information of other columns within the same table, respectively. Due to space limitations, some textual rules have been omitted. The complete prompt template is provided in the supplementary material.

\noindent\textbf{Prompting and results.} The prompt can be generated by filling in the prompt template slots with information from the target column, after which the prompt is input into the LLMs to obtain the results. The basic principle of LLM inference involves autoregressively generating the next token with the highest probability based on the input prompt, continuing until an end-of-sentence marker is produced. We can derive the final estimation and the reason from the sentence generated by the LLM.

\subsection{Experiments and Discussion}\label{sec:exp:llm}
\noindent\textbf{LLM setup.} We employ several SOTA open-source and commercial LLMs to investigate their performance in the task of NDV estimation. Specifically, we use the instruct version of Llama 3.1~\cite{llama31} with 8B and 70B parameters, denoted as \texttt{Llama31-8B} and \texttt{Llama31-70B}; the instruct version of Qwen2~\cite{yang2024qwen2} 7B and 72B parameters, denoted as \texttt{Qwen2-7B} and \texttt{Qwen2-72B}; as well as OpenAI \texttt{GPT3.5}(gpt-35-turbo-0125)~\cite{gpt35} and \texttt{GPT4o} (gpt-4o-2024-05-13)~\cite{gpt4o}. The prompt of each LLM is using the template in Listing~\ref{lst:method:column} filled with information of each column. Due to the cost issues associated with invoking LLMs, we conduct experiments solely in the limited sequential data access scenarios.

\begin{table}[t]
    \centering
    \caption{Performance of prompting LLMs under sequential access 100 rows condition.}
    \begin{tabular}{ccccccc}
\toprule
         & Mean & 50\% & 75\%& 90\% & 95\% & 99\% \\
\midrule
        \texttt{Llama31-8B} & 3.5e6 & 6.15 & 82.81 & 602.12 & 1.87e3 & 1.29e4 \\   
        \texttt{Qwen2-7B} & 2.5e5 & 9.14 & 102.55 & 847.39 & 4.00e3 & 4.45e4 \\          
        \texttt{Llama31-70B} & 213.8 & 2.00 & 9.11 & 59.63 & 191.85 & 1.73e3 \\       
        \texttt{Qwen2-72B} & 230.6 & 2.16 & 10.85 & 74.0 & 232.19 & 2.64e3 \\         
\midrule
        \texttt{GPT3.5} & 5e14 & 4.17 & 31.04 & 170.28 & 515.26 & 1.02e4 \\ 
        \texttt{GPT4o} & 79.25 & 2.00 & 6.00 & 28.53 & 90.73 & 1.09e3 \\ 
\midrule
        \texttt{PLM4NDV} & 13.33 & 1.86 & 3.81 & 10.81 & 25.18 & 148.76 \\ 
\bottomrule
    \end{tabular}
    \label{tab:llm}
    \vspace{-3mm}
\end{table}

\noindent\textbf{Overall results.} The overall performance of the LLMs is presented in Table~\ref{tab:llm}. Based on the results in Table~\ref{tab:llm} and the results from sequential access conditions reported in Table~\ref{tab:mainexp} and Table~\ref{tab:mainexp2}, we can derive the following conclusions:
\begin{itemize}[leftmargin=15pt]
    \item LLMs are applicable to NDV estimation. By using prompting methods, the NDV estimation task can be transformed into a natural language question, allowing for the estimation to be accomplished by leveraging the capabilities of LLMs.
    \item Among the six LLMs, \texttt{GPT4o} demonstrates the best overall performance. Additionally, the larger-sized LLMs (\texttt{Llama31-70B} and \texttt{Qwen2-72B}) perform significantly better than their smaller counterparts (\texttt{Llama31-8B} and \texttt{Qwen2-7B}) for the same architecture. The overall performance of NDV estimation using LLMs is closely related to their semantic understanding and reasoning capabilities.     
    \item The performance of prompting \texttt{GPT4o} surpasses most baselines in most metrics. Particularly, the 75\% percentile of q-error is 6.00, demonstrating its superior accuracy compared to the baselines in most test cases. In addition, \texttt{GPT4o} and \texttt{Llama31-70B} outperform all the baseline methods at the 50\% percentile of q-error. This suggests that prompting LLMs may derive outstanding results in considerable test cases, however, poor estimation outcomes in a few cases lead to overall suboptimal performance.
    \item For the LLMs having poor overall performance (\texttt{Llama31-8B}, \texttt{Qwen2-7B}, and \texttt{GPT3.5}), the mean error is significantly higher than the 99\% percentile. In addition, the mean q-error is greater than the 90\% percentile for all LLMs, indicating that their estimation errors are substantial in extreme scenarios, contributing to this phenomenon. 
    \item \texttt{PLM4NDV} outperforms all the LLMs across all metrics. This can be attributed to the fact that \texttt{PLM4NDV} is trained for the NDV estimation task using the training datasets, whereas the LLMs are not. Fine-tuning LLMs for the NDV estimation task using existing data with schema, sample records, and ground truth NDV is a non-trivial task, as there are no intermediate results to map them together. The underlining rules can be very complex as shown in Equation~(\ref{eq:sichel}) and Equation~(\ref{eq:goodman}). \texttt{PLM4NDV} learns this mapping from the training data, incorporating semantic information, which contributes to its superior performance.
\end{itemize}

\begin{lstlisting}[style=prompt, label=lst:exp:case, caption={A case study of LLM for NDV estimation}, float=t]
{
  "Reason": "The column name 'batting-position' suggests that it represents the batting position of a player in a sports game, likely cricket or baseball. In these sports, batting positions typically range from 1 to 11. However, considering the total rows (37963), it is unlikely that there are 37963 unique batting positions. A reasonable estimate would be around 1000 distinct batting positions, assuming there are multiple players and games represented in the dataset.",
  "NDV": "1000"  
}
\end{lstlisting}

\noindent\textbf{Case study.} To intuitively demonstrate how we derive the estimation $\hat{D}$ from the LLM output and how the LLM performs reasoning based on the prompts, we present an output case in Listing~\ref{lst:exp:case}. From the JSON format output in the listing, we can obtain the estimation $\hat{D}=1000$ along with the reasoning process behind the estimation.

By examining the reasons in the LLM output, we find that, on the one hand, LLM can infer hidden semantic information and specific meanings based solely on the input column names, even identifying the ground-truth NDV range (from 1 to 11). On the other hand, providing the total row count is necessary for the LLM to address the NDV estimation problem by answering the natural language question. However, due to potential limitations in the reasoning ability of the LLM itself or the suboptimal engineering of the prompt template, the LLM may understand the semantics of the column but still fail to present an accurate estimation.

\noindent{\textbf{Lessons.}} From the above observations, we can derive the following key insights.

\begin{itemize}[leftmargin=15pt]
    \item LLMs show great potential in solving NDV estimation tasks, as evidenced by the performance of three LLMs (\texttt{Llama31-70B}, \texttt{Qwen2-72B}, and \texttt{GPT4o}), which achieve a q-error of at most 2.16 in the 50th percentile. The performance of the three LLMs beat most baselines in Table~\ref{tab:mainexp} in the 50th percentile.
    \item The case study directly demonstrates the benefits of semantic information. The reasons provided in the LLM output indicate that the LLM can extract semantic information from the schema, which aids in NDV estimation. It can infer the meaning and the value domain based on the column name.
    \item However, prompting LLMs for NDV estimation may limited by their reasoning abilities, as their estimations are very poor in a few scenarios. LLMs may understand the meaning of the column but is difficult to estimate NDV, as shown in the case study.
    \item Fine-tuning LLMs for NDV estimation is challenging due to the lack of intermediate results linking the schema, sample records, and ground truth NDV. This makes it difficult to train the model to understand these relationships, requiring careful structuring of the training data.
    \item Therefore, incorporating LLMs with existing NDV estimation techniques may be a challenging and interesting topic for future research. 
\end{itemize}

\section{Related Work}\label{sec:relatedwork}
Our work is positioned at the intersection of NDV estimation and the application of PLMs/LLMs.
\subsection{NDV Estimation}

\noindent\textbf{Sketch-based methods.} Sketch-based methods need to access the entire data column to maintain a memory-efficient sketch and then estimate the NDV from the sketch~\cite{durand2003loglog,mincount_giroire2009order,flajolet2007hyperloglog,whang1990linear}. Sketch-based methods can be classified into two subcategories. \textit{Logarithmic hashing} methods use a bitmap and a hash function to track the most uncommon element observed so far, and the hash function maps each element with a probability varying by the bit significance. The representative methods in this subcategory are \texttt{FM}~\cite{flajolet1985probabilistic}, \texttt{PCSA}~\cite{flajolet1985probabilistic}, \texttt{AMS}~\cite{alon1996space}, \texttt{LogLog}~\cite{durand2003loglog}, SuperLogLog~\cite{durand2003loglog}, HyperLogLog~\cite{flajolet2007hyperloglog},  HyperLogLog++~\cite{hllpp_heule2013hyperloglog}, and UltraLogLog~\cite{ertl2023ultraloglog}. \textit{Uniform hashing} methods use a uniform hash function to hash the entire dataset into an \textit{interval} or a set of \textit{buckets}. The former is based on how packed the interval is, and the representative methods are \texttt{Cohen}~\cite{cohen1997size}, \texttt{BJKST1}~\cite{bjkst1_bar2002counting}, \texttt{AKMV}~\cite{akmv_beyer2007synopses}, and \texttt{MinCount}~\cite{mincount_giroire2009order}. The latter is based on the probability that a bucket is (non)empty, and the typical methods are \texttt{BJKST2}~\cite{bjkst1_bar2002counting}, Linear Counting~\cite{whang1990linear}, and Bloom Filter~\cite{papapetrou2010cardinality}.

\noindent\textbf{Sampling-based methods.} Since sampling-based methods utilize a portion of IID samples rather than the full data for estimation, most of these methods are based on various heuristics formulated from specific assumptions~\cite{goodman1949estimation}. For instance, \texttt{Chao}~\cite{chao1984nonparametric,chao_in_db_ozsoyoglu1991estimating} and \texttt{ChaoLee}~\cite{chaolee} assume the size of the column is infinite, \texttt{Shlosser}~\cite{shlosser1981estimation} supposes there are certain conditions of data skewness, Horvitz-Thompson (\texttt{HT})~\cite{horvitz_sarndal1992model} and Bootstrap~\cite{bootstrap_smith1984nonparametric} hypothesize the distinct elements in the sample are distinct in the full dataset, Method of Movement (\texttt{MoM1})~\cite{mmo_bunge1993estimating} postulates the column size is infinite and the frequency of each distinct element is constant, the variant of \texttt{MoM1} (\texttt{MoM2})~\cite{mmo_bunge1993estimating} removes the infinite size assumption, and another variant of \texttt{MoM1} (\texttt{MoM3})~\cite{mmo_bunge1993estimating} assumes the frequency of each distinct element are unequal. \texttt{GEE}~\cite{gee_charikar2000towards} is constructed to match the theoretical lower bound of estimation error. The Error Bound (\texttt{EB})~\cite{error_bound} estimator takes the estimation in the sampling process and claims its estimation error can be bounded. \texttt{Sichel}~\cite{sichel1986parameter,sichel1986word,sichel1992anatomy} aims to fit a zero-truncated generalized inverse Gaussian-Poisson distribution to estimate NDV. 
\texttt{Jackknife}~\cite{burnham1978estimation,burnham1979robust} and Smoothed Jackknife (\texttt{SJ})~\cite{hybskew_haas1995sampling} assume the existence of a non-zero constant that can satisfy an expression to correct the estimation error iteratively. Such harsh assumptions can hardly be satisfied in practice, resulting in large estimation errors. In recent years, Machine Learning (ML) techniques~\cite{apml_acharya2017unified,pml_orlitsky2004modeling,apml_pavlichin2019approximate,hao2019unified,wu2019chebyshev} have been introduced in NDV estimation~\cite{ls_wu2022learning,li2024learning,xu2025adandv}. 
Besides, numerous studies have made much effort to improve the efficiency of sampling in databases~\cite{error_bound,chaudhuri2004effective,doucet2006efficient,chen2022efficient,sanca2023laqy,ci2015efficient}.

\subsection{Pre-trained Language Models}
PLMs~\cite{kenton2019bert,gpt_radford2018improving,gpt2_radford2019language} have achieved SOTA results in various NLP tasks. In recent years, large-scale PLMs, known as Large Language Models (LLMs), have emerged, characterized by their extensive parameters and remarkable advancements. Refer to~\cite{han2021pre,zhao2023survey} for a comprehensive review, in this paper, we focus on their applications in databases.

\noindent\textbf{PLMs in Databases.} The most representative application of PLMs in databases is Text-to-SQL, for which a comprehensive literature review is provided in~\cite{qin2022survey}. Additionally, the database community has integrated PLMs into various data management and database tasks. For instance, Li et al.~\cite{li2020deep} use PLMs as a base model for entity matching, Deng et al.~\cite{deng2022turl} utilize PLMs as pre-training frameworks for table understanding, and Suhara et al.~\cite{suhara2022annotating} leverage PLMs to annotate columns in databases.

\noindent\textbf{LLMs in Databases.} Numerous recent works in Text-to-SQL based on LLM have emerged~\cite{gao2023text,gu2023few,gu2023interleaving,pourreza2024din}. Leveraging the advantages of LLMs in few-shot and zero-shot performance, these works have led to innovative approaches in database management. For instance, FormaT5~\cite{singh2023format5} utilizes LLM for table formatting, \textsc{GenRewrite}~\cite{liu2024query} uses LLM to rewrite SQL for performance improvement, and LLMTune~\cite{huang2024llmtune} accelerates the knob tuning process using LLMs. Furthermore, recent efforts have attempted to build holistic LLM-centric database optimization and diagnosis systems~\cite{zhou2023llm,zhou2024dbot,zhou2024llm}.

\section{Conclusion}\label{sec:conclusion}
In this paper, we aim to minimize data access in NDV estimation and present several distinct contributions, marking a significant advancement in the field. We propose \texttt{PLM4NDV}, the pioneer NDV estimation method that leverages schema information using Pre-trained Language Models. We are the first to highlight the benefits of the historically ignored schema information in NDV estimation. By integrating the schema semantics, our method substantially augments the accuracy of NDV estimation under limited sampling cost budgets. Besides, we show that \texttt{PLM4NDV} can establish the estimation even without data access and achieve promising performance. Consequently, we provide a third NDV estimation paradigm that differs from existing sketch-based and sampling-based approaches. Our extensive empirical studies on a large-scale real-world dataset demonstrate that \texttt{PLM4NDV} significantly reduces the data access and improves NDV estimation accuracy compared to baselines.
 
\textbf{Future work}. Our work showcases the promising new opportunities in NDV estimation and raises several open research questions. For instance, how to serialize column schemas to obtain more effective semantic embeddings, generate the missing optional components in the schema and leverage them, discern the impact of data types, fine-tune PLMs in the context of database schema, and effectively leverage LLMs for NDV estimation. We look forward to exploring these questions with the community in the future.

\bibliographystyle{ACM-Reference-Format}
\bibliography{main}

\end{document}